\documentclass[12pt]{article}
\usepackage{axodraw}
\usepackage{graphicx}
\usepackage{amsmath}

\usepackage{xcolor}
\usepackage[affil-it]{authblk}

\newcommand{\sla}[1]{#1\!\!\!\slash}

\newcommand{\re}{\mathop{\mathrm{Re}}\nolimits}
\newcommand{\li}{\mathop{\mathrm{Li}}\nolimits}

\allowdisplaybreaks

\begin{document}

\title{
\vskip-3cm{\baselineskip14pt
\centerline{\normalsize DESY 15-032\hfill ISSN 0418-9833}
\centerline{\normalsize March 2015\hfill}}
\vskip1.5cm
Two-loop electroweak threshold corrections in the Standard Model}

\author[1]{Bernd A. Kniehl\thanks{kniehl@desy.de}}
\author[1,2]{Andrey F. Pikelner\thanks{pikelner@theor.jinr.ru}}
\author[1]{Oleg L. Veretin\thanks{veretin@mail.desy.de}}

\affil[1]{\rm
  II. Institut f\"ur Theoretische Physik, 
  Universit\"at Hamburg, \authorcr  
  Luruper Chaussee 149, 22761 Hamburg, Germany}

\affil[2]{\rm
  Bogoliubov Laboratory of Theoretical Physics, 
  Joint Institute for Nuclear Research, \authorcr 
  141980 Dubna, Russia}

\date{}

\maketitle

\begin{abstract}
We study the relationships between the basic parameters of the on-shell
renormalization scheme and their counterparts in the $\overline{\mathrm{MS}}$
scheme at full order ${\cal O}(\alpha^2)$ in the Standard Model.
These enter as threshold corrections the renormalization group analyses
underlying, e.g., the investigation of the vacuum stability.
To ensure the gauge invariance of the parameters, in particular of the
$\overline{\mathrm{MS}}$ masses, we work in $R_\xi$ gauge and systematically
include tadpole contributions.
We also consider the gaugeless-limit approximation and compare it with the
full two-loop electroweak calculation. 

\medskip

\noindent
PACS numbers: 11.10.Gh, 11.10.Hi, 12.15.Lk, 14.80.Bn

\end{abstract}

\newpage

\section{Introduction}
\label{sec:introduction}

The measurement of the Higgs boson mass at the Large Hadron Collider
\cite{Aad:2012tfa} not only fully confirms the validity of the Standard Model
(SM) around the electroweak scale, but also opens a possibility for a precise
study of the applicability of the SM at energies of the order of the Planck
mass.
Renormalization group (RG) equations, which determine the dependence on the
renormalization scale $\mu$ of the running parameters, which are usually
defined in the modified minimal-subtraction ($\overline{\mathrm{MS}}$) scheme
of dimensional regularization, play an essential r\^ole in such analyses.
In the SM, the corresponding RG functions, whose knowledge had been limited to
the two-loop order for a very long time \cite{Fischler:1982du}, have recently
been evaluated at three loops for all running parameters, including gauge,
Yukawa, and scalar self couplings \cite{Mihaila:2012fm}.

The other aspect of the problem is the matching between the
$\overline{\mathrm{MS}}$ parameters and the physical observables, which gives
rise to threshold corrections.\footnote{%
The usage of the term {\it threshold corrections} in this context is to
indicate that the initial conditions for the RG evolution of the
$\overline{\mathrm{MS}}$ parameters are determined at some low-lying scale.
This term also appears in the effective-field-theory language, where it
carries a different meaning.}
These not only include terms of the form $\ln\mu^2$, but also non-logarithmic
ones.
The relationships between the $\overline{\mathrm{MS}}$ and pole masses of the
intermediate bosons were obtained at the two-loop level in
Refs.~\cite{Jegerlehner:2001fb,Jegerlehner:2002em}.
As for the threshold corrections to the top and bottom quark masses and Yukawa
couplings, the situation is as follows.
The QCD corrections, which are dominant, are available at one
\cite{Braaten:1980yq}, two \cite{Gray:1990yh,Fleischer:1998dw}, three
\cite{Chetyrkin:1999ys}, and four \cite{Marquard:2015qpa} loops. 
The two-loop result in the supersymmetric extension of QCD was obtained in
Ref.~\cite{Bednyakov:2002sf}.
The one-loop electroweak corrections, of order $O(\alpha)$, were first
considered in Ref.~\cite{Hempfling:1994ar}.
The two-loop mixed $O(\alpha\alpha_s)$ corrections were provided for the bottom
quark in Ref.~\cite{Kniehl:2004hfa} and for the top quark in
Ref.~\cite{Jegerlehner:2003py,Jegerlehner:2012kn,Bezrukov:2012sa}.
Recently, also the two-loop electroweak corrections of order $O(\alpha^2)$ have
been obtained in the gaugeless-limit approximation \cite{Kniehl:2014yia}.
Also the threshold corrections to the self-coupling constant of the scalar
field were intensively studied in the literature.
The $O(\alpha)$ corrections were evaluated a long time ago in
Ref.~\cite{Sirlin:1985ux} and the $O(\alpha\alpha_s)$ ones recently in
Ref.~\cite{Bezrukov:2012sa}.
As for the $O(\alpha^2)$ corrections, the leading term was found in
Ref.~\cite{Degrassi:2012ry}, and an interpolation formula, which also includes
subleading contributions, was given in Ref.~\cite{Buttazzo:2013uya}.
These analyses were recently revisited in Ref.~\cite{Martin:2014cxa} by
providing precise numerical results.

In this paper, we systematically present the complete two-loop threshold
corrections, from the orders $O(\alpha)$, $O(\alpha\alpha_s)$, and
$O(\alpha^2)$, to all the running parameters of the SM independently obtained
by an analytic calculation.
This includes the masses of the $W$, $Z$, and Higgs bosons ($m_W,m_Z,m_H$) and
those of the top and bottom quarks ($m_t,m_b$) as well as the gauge couplings
($g,g^\prime$), the Higgs self-coupling ($\lambda$), and the top and bottom
Yukawa couplings ($y_t,y_b$).
In contrast to Refs.~\cite{Buttazzo:2013uya,Martin:2014cxa,Martin:2013gka},
all our calculations are performed in $R_\xi$ gauge keeping the gauge-fixing
parameters free, which allows us to explicitly track the $\xi$ dependencies and
so to ensure the gauge independence of the threshold corrections and the
$\overline{\mathrm{MS}}$ parameters.
The tadpole diagrams turn out to play a crucial r\^ole in this
(see Subsection~\ref{sec:tadpoles}).

This paper is organized as follows.
In Section~\ref{sec:setup}, we set up the stage for our calculation of the
threshold corrections. 
In Subsections~\ref{sec:deltar}--\ref{sec:alpha}, we discuss the various
ingredients entering our analysis.
Our results are presented in Section~\ref{sec:results} and
Appendix~\ref{app:res}.
In Appendix~\ref{app:zmmH}, we also list the $\overline{\mathrm{MS}}$
renormalization constant of the Higgs boson mass.


\section{Setup}
\label{sec:setup}

The SM may exist in two different phases: the symmetric phase and the phase
with the spontaneously broken symmetry.
The phase is determined by the potential of the
scalar field $\phi$,
\begin{equation}
  V(\phi) = m_\phi^2 \,\phi^\dagger\phi + \lambda\, (\phi^\dagger \phi)^2 \,,
\label{potential}  
\end{equation}
where $m_\phi$ is a mass parameter and $\lambda$ is the self-coupling constant
of the scalar field.
While stability requires $\lambda>0$, the term $m_\phi^2$ can be either
positive (symmetric phase) or negative (broken phase).
In the symmetric phase, the SM is naturally parametrized by the following set
of parameters:
\begin{equation}
  g, g', \lambda, m_\phi, y_f.
\label{set1}  
\end{equation}
In the broken phase, it is convenient to choose an alternative set of
parameters, namely
\begin{equation}
  e, m_W, m_Z, m_H, m_f,
\label{set2}  
\end{equation}
where $e$ is the electromagnetic gauge coupling.

At the tree level, the parameters in Eqs.~(\ref{set1}) and (\ref{set2}) can be
related to each other by
\begin{equation}
 \frac{1}{e^2} = \frac{1}{g^2} + \frac{1}{g'^2}
\label{ee}
\end{equation}
for the couplings and
\begin{equation}
  \frac{4m_W^2}{v^2} = g^2 \,, \qquad
  \frac{4m_Z^2}{v^2} = g^2 + g'^2 \,, \qquad
  \frac{m_H^2}{2v^2} = \lambda \,, \qquad
  \frac{2m_f^2}{v^2} = y_f^2\,,
\label{mm}
\end{equation}
for the masses, where we have introduced the vacuum expectation value of the
scalar field $v\approx246$~GeV, which characterizes the broken phase.
This parameter is not independent, since the parameter sets in
Eqs.~(\ref{set1}) and (\ref{set2}) each fully determine the
theory.\footnote{%
Throughout this paper, we take the quark mixing matrix to be unity.}
In fact, we have
\begin{equation}
 v=\sqrt{\frac{-m_\phi^2}{\lambda}}
\end{equation}
in terms of Eq.~(\ref{set1}) and
\begin{equation}
\frac{1}{v^2}= \frac{e^2}{4m_W^2(1-m_W^2/m_Z^2)}
\label{vv}
\end{equation}
in terms of Eq.~(\ref{set2}).

Equations~(\ref{ee})--(\ref{vv}) are subject to radiative corrections.
However, we have the freedom to choose 16 out of the 32 parameters in
Eqs.~(\ref{set1}) and (\ref{set2}) to be ``observables" and to consider the
remaining ones to be derived parameters of the theory. 
At energies of the order of $v$, the parameters in Eq.~(\ref{set2}) are
intuitively closer to what we would call ``observables,'' especially if we
define them in the on-shell renormalization scheme.
Then, these include Sommerfeld's fine-structure constant
$\alpha_{\mathrm{Th}}=e_{\mathrm{Th}}^2/(4\pi)$ as measured in Thomson
scattering and the pole masses $M_W$, $M_Z$, $M_H$, and $M_f$, which are the
zeroes of the inverses of the propagators of the respective
particles.
By contrast, the parameters in Eq.~(\ref{set1}) are more suitable for studies
of the RG evolution.
In this work, we shall discuss the relationships between these parameters and
the radiative corrections to Eqs.~(\ref{ee})--(\ref{vv}).

As usual, we consider Eqs.~(\ref{ee})--(\ref{vv}) to be valid through all
orders of perturbation theory by definition.
For the reason explained above, we define the parameters in Eq.~(\ref{set1}),
which appear on the right-hand sides of Eqs.~(\ref{ee}) and (\ref{mm}), in
the $\overline{\mathrm{MS}}$ renormalization scheme, which then carries over
to the parameters in Eq.~(\ref{set2}), which appear on the left-hand sides of
Eqs.~(\ref{ee}) and (\ref{mm})
(see Subsections~\ref{sec:masses} and \ref{sec:alpha}).
The $\mu$ dependencies of the parameters in Eq.~(\ref{set1}) are governed by the
RG equations.
These allow us to run the parameters from a few GeV way up to the Planck mass.
The relevant $\beta$ functions are available at two \cite{Fischler:1982du} and
three loops \cite{Mihaila:2012fm}.
The masses in Eq.~(\ref{set2}) also obey RG equations.
The corresponding RG functions are available at the two-loop order from
Refs.~\cite{Jegerlehner:2001fb,Jegerlehner:2002em,Kniehl:2014yia} and are
partly recovered here by an independent calculation.
As explained in Refs.~\cite{Jegerlehner:2001fb,Jegerlehner:2002em}, we can
always relate the RG functions of the unbroken and broken phases, which serves
as a welcome check for the correctness of our results.

As for the initial conditions for the RG evolution, the
$\overline{\mathrm{MS}}$ parameters are usually expressed in terms of the
parameters of the on-shell renormalization scheme,
$\alpha_{\mathrm{Th}}$, $M_W$, $M_Z$, $M_H$, and $M_f$.
The matching scale where this is done is typically chosen to be of the order of
$M_Z$ or $M_t$.
In this case, one may safely neglect the masses of all fermions, except for
the one of the top quark and possibly also the one the bottom quark. 
We shall retain the full dependence on the bottom-quark mass at one loop, but
neglect it at two loops.
The strong-coupling constant $\alpha_s(\mu)$ is always defined in the
$\overline{\mathrm{MS}}$ renormalization scheme.
Finally, $v$ is related to $e$ via Eq.~(\ref{vv}).
Phenomenologically, it is more convenient to express the
$\overline{\mathrm{MS}}$ parameter $v(\mu)$ in terms of the Fermi constant
$G_F$, which is measured in low-energy processes of the weak interaction, such
as muon decay, through the exact relationship
\begin{equation}
2^{1/2} G_F=\frac{1+\Delta\overline{r}(\mu)}{v^2(\mu)}\,,
\label{dRdef0}
\end{equation}
where $\Delta\overline{r}(\mu)$ is an appropriate variant of Sirlin's
$\Delta r$ parameter \cite{Sirlin:1980nh} (see Subsection~\ref{sec:deltar}).
Inserting Eq.~(\ref{dRdef0}) in Eq.~(\ref{mm}) and accommodating the
relationships between the $\overline{\mathrm{MS}}$ and pole masses, one may
cast the relations of interest in the form
\begin{eqnarray}
  g^2(\mu) &=& 2^{5/2} G_F M_W^2 [ 1 + \delta_W(\mu) ] \,,
\label{g2}
\\
  g^2(\mu) + g'^2(\mu) &=& 2^{5/2} G_F M_Z^2 [ 1 + \delta_Z(\mu) ] \,,
\label{g2g2}
\\
  \lambda(\mu) &=& 2^{-1/2} G_F M_H^2 [ 1 + \delta_H(\mu) ] \,,
\label{lambda}
\\
  y_f(\mu) &=& 2^{3/4} G_F^{1/2} M_f [ 1 + \delta_f(\mu) ] \,.
\label{y2}
\end{eqnarray}
In turn, substituting Eq.~(\ref{dRdef0}) in the right-hand sides of
Eqs.~(\ref{g2})-(\ref{y2}) and equating the outcome with the right-hand sides
of Eq.~(\ref{mm}), we may write the $\overline{\mathrm{MS}}$ to pole mass
relationships as
\begin{eqnarray}
  m_x^2(\mu) &=& M_x^2 [1+\Delta\overline{r}(\mu)] [ 1 + \delta_x(\mu) ],
      \qquad x=W,Z,H\,,
\nonumber\\
  m_x(\mu) &=& M_x [1+\Delta\overline{r}(\mu)]^{1/2} [ 1 + \delta_x(\mu) ],
      \qquad x=t,b\,,
\label{mfer}
\end{eqnarray}
which are valid to all orders of perturbation theory.
It is the goal of this work to evaluate $\Delta\overline{r}(\mu)$ and $\delta_x(\mu)$
for $x=W,Z,H,t,b$ through order $O(\alpha^2)$.


\boldmath
\subsection{Vacuum expectation value and $\Delta\overline{r}(\mu)$}
\label{sec:deltar}
\unboldmath

We start our discussion with the evaluation of the threshold corrections to
$v(\mu)$, i.e., of $\Delta\overline{r}(\mu)$ in Eq.~(\ref{dRdef0}).
As already mentioned above, this quantity is very similar to $\Delta r$ in the
on-shell scheme \cite{Sirlin:1980nh}, which contains the radiative corrections
to the muon lifetime that the SM introduces on top of the electromagnetic
corrections in the Fermi model of four-fermion interactions.
The difference between $\Delta r$ and $\Delta\overline{r}(\mu)$ is that the former is
defined in the on-shell scheme according to 
\begin{equation}
G_F=\frac{\pi\alpha_\mathrm{Th}}{\sqrt{2}M_W^2(1-M_W^2/M_Z^2)}(1+\Delta r)\,,
\label{drdef}
\end{equation}
while the latter obeys the analogous relation with the on-shell parameters
replaced by their $\overline{\mathrm{MS}}$ counterparts,
\begin{eqnarray}
G_F=\frac{\pi\alpha(\mu)}{\sqrt{2}m_W^2(\mu)[1-m_W^2(\mu)/m_Z^2(\mu)]}
[1+\Delta\overline{r}(\mu)] \,.
\label{dRdef}
\end{eqnarray}
Consequently, the calculations of $\Delta r$ and $\Delta\overline{r}(\mu)$ are very
similar.
In particular, they are both based on the matching of the Fermi model and the
SM and exhibit a factorization of the low-energy scales.
The contribution of the light (massless) fermions to $\Delta\overline{r}(\mu)$ may be
found in Ref.~\cite{Malde:1999bd}.
The $\overline{\mathrm{MS}}$ parameter $\alpha(\mu)$ deserves a detailed
discussion, which will be presented in Subsection~\ref{sec:alpha}.

The framework for the evaluation of $\Delta r$ or $\Delta\overline{r}(\mu)$ at any loop
order was established in Refs.~\cite{Awramik:2002vu,Onishchenko:2002ve}. 
One may start from any amplitude involving the exchange of the weak charged
current in the SM, e.g., $A(e+\nu_e\to\mu+\nu_\mu)$, where
$e$, $\nu_e$, $\mu$, and $\nu_\mu$
are the electron, the electron neutrino, the muon, and the muon neutrino,
respectively.
As demonstrated in Ref.~\cite{Awramik:2002vu}, there is a factorization theorem
that allows for the convenient separation of the soft scales
(see also the discussion in Ref.~\cite{Actis:2006rc}). 
Applying this to the evaluation of $\Delta\overline{r}(\mu)$, we only need to perform the
renormalization procedure in the $\overline{\mathrm{MS}}$ scheme.
According to Ref.~\cite{Awramik:2002vu}, we may write
\begin{eqnarray}
\frac{e^2}{8m_W^2(1-m_W^2/m_Z^2)}( 1+\Delta\overline{r} )
     = \left[ \sqrt{Z_{2,e}Z_{2,\nu_e}Z_{2,\mu}Z_{2,\nu_\mu}} 
               A(e+\nu_e\to\mu+\nu_\mu)   \right]_\mathrm{hard}  \,,
\label{dr}
\end{eqnarray}
where $Z_{2,f}$ is the wave function renormalization constant of the
left-handed field component of fermion $f$ in the $\overline{\mathrm{MS}}$
scheme.
The subscript {\it hard} in Eq.~(\ref{dr}) implies that all external
four-momenta and all the light-fermion masses are identically put to zero
{\it before} the integration over the loop momenta.\footnote{%
In other words, all the loop momenta $k_i$ are {\it hard} compared to the
low-energy scales, such as the muon mass $m_\mu$, that is $|k_i|\gg m_\mu$.}
Obviously, such a procedure generates a lot of infrared divergences in the
calculation.
These are regularized by the dimensional-regularization
parameter $\varepsilon$ and cancel on the right-hand side of Eq.~(\ref{dr}).
This cancellation is nontrivial, and this is exactly the statement of the
factorization theorem \cite{Awramik:2002vu}.
Following this procedure, the evaluation of $\Delta\overline{r}$ is completely
reduced to vacuum diagrams with one or two loops.
In Appendix~\ref{app:res}, we shall present an analytical expression for
$\Delta\overline{r}$ in terms of $\overline{\mathrm{MS}}$ couplings and pole masses.


\subsection{Role of tadpoles}
\label{sec:tadpoles}

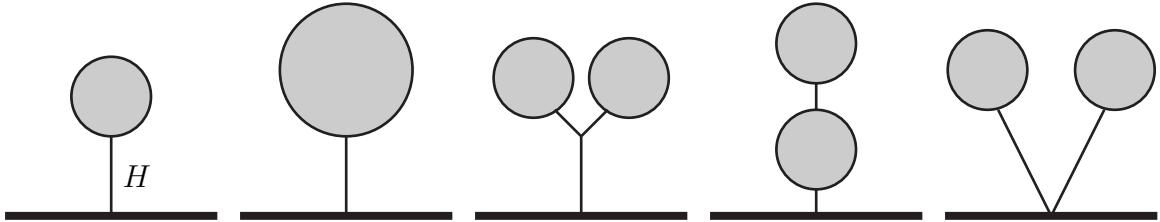
\begin{figure}[ht]
\centerline{
\begin{picture}(85,80)(0,0)
\SetScale{1.0}
\GCirc(40,45){15}{0.8}
\SetWidth{3.0}
\Line(0,0)(80,0)
\SetWidth{1.0}
\Line(40,0)(40,30)
\CArc(40,45)(15,0,360)
\Text(50,15)[c]{$H$}
\end{picture}
\begin{picture}(85,80)(0,0)
\SetScale{1.0}
\GCirc(40,55){25}{0.8}
\SetWidth{3.0}
\Line(0,0)(80,0)
\SetWidth{1.0}
\Line(40,0)(40,30)
\CArc(40,55)(25,0,360)
\end{picture}
\begin{picture}(85,80)(0,0)
\SetScale{1.0}
\GCirc(22,52){15}{0.8}
\GCirc(58,52){15}{0.8}
\SetWidth{3.0}
\Line(0,0)(80,0)
\SetWidth{1.0}
\Line(40,0)(40,30)
\Line(40,30)(30,40)
\Line(40,30)(50,40)
\CArc(22,52)(15,0,360)
\CArc(58,52)(15,0,360)
\end{picture}
\begin{picture}(85,80)(0,0)
\SetScale{1.0}
\GCirc(40,25){15}{0.8}
\GCirc(40,65){15}{0.8}
\SetWidth{3.0}
\Line(0,0)(80,0)
\SetWidth{1.0}
\Line(40,0)(40,10)
\Line(40,40)(40,50)
\CArc(40,25)(15,0,360)
\CArc(40,65)(15,0,360)
\end{picture}
\begin{picture}(85,80)(0,0)
\SetScale{1.0}
\GCirc(16,55){15}{0.8}
\GCirc(64,55){15}{0.8}
\SetWidth{3.0}
\Line(0,0)(80,0)
\SetWidth{1.0}
\Line(40,0)(20,40)
\Line(40,0)(60,40)
\CArc(16,55)(15,0,360)
\CArc(64,55)(15,0,360)
\end{picture}
}
\caption{\label{fig:one}%
One- and two-loop tadpole contribution to the propagator of a particle.
$H$ stands for the Higgs boson propagator with zero momentum transfer.}
\end{figure}

An important comment is in order here.
In a theory with a broken gauge symmetry, it is important to take into account
the tadpole diagrams.
One- and two-loop tadpole diagrams contributing to the propagator of a particle
are shown in Fig.~\ref{fig:one}.
All such tadpole insertions must be made not only in the Feynman diagrams
contributing to the counterterms, but also in all the proper Feynman diagrams
to ensure the gauge independence of renormalized scattering amplitudes.
In particular, the mass counterterms are gauge dependent unless the tadpole
contributions are included, as has been known for a long time
\cite{Fleischer:1980ub}.
This is also true for the threshold corrections, as was first observed for the
one-loop electroweak threshold corrections to the Yukawa
\cite{Hempfling:1994ar} and Higgs self-couplings \cite{Sirlin:1985ux}.

In this work, we perform all calculations at two loops in $R_\xi$ gauge with
four independent gauge parameters, related to the $W$ and $Z$ bosons, the
photon, and the gluon.
We verify by explicit calculation that the threshold corrections $\Delta\overline{r}$ in
Eq.~(\ref{dRdef0}) and $\delta_x$ in Eqs.~(\ref{g2})--(\ref{y2}) are gauge
independent.
This serves as strong check for the correctness of our results.

Tadpole contributions are singular in the limit $M_H\to0$.
The most divergent terms scale as $1/M_H^2$ at one loop and as $1/M_H^4$ at two
loops.
This behavior is fully inherited by $\Delta\overline{r}$ and the
$\overline{\mathrm{MS}}$ masses.
However, nontrivial cancellations between $\Delta\overline{r}$ and the
$\overline{\mathrm{MS}}$ masses take place in the threshold corrections
$\delta_W$, $\delta_Z$, $\delta_t$, and $\delta_b$, which render the latter
finite in the limit $M_H\to0$, i.e., also terms proportional to
$M_H^0\ln M_H^2$ are canceled.
As for the Yukawa couplings, this was noticed at $\mathcal{O}(\alpha)$ in
Ref.~\cite{Hempfling:1994ar} and at $\mathcal{O}(\alpha\alpha_s)$ in
Ref.~\cite{Kniehl:2004hfa} for the bottom quark and in
Ref.~\cite{Bezrukov:2012sa} for the top quark.
Here, we investigate the behavior for $M_H\to0$ at $\mathcal{O}(\alpha^2)$ and
find that $\Delta\overline{r}$ and the $\overline{\mathrm{MS}}$ masses contain terms
proportional to $1/M_H^4$, while $\delta_W$, $\delta_Z$, $\delta_t$, and
$\delta_b$ are finite in this limit.

The situation is different in the Higgs sector.
Already at one loop, the threshold corrections to $m_H$ and $\lambda$ contain
terms that diverge as $1/M_H^2$ for $M_H\to0$ \cite{Sirlin:1985ux}.
At two loops, however, the $1/M_H^4$ terms cancel, so that the leading
small-$M_H$ behaviors go unchanged.
This cancellation ensures that $\lambda$ stays finite in the limit $M_H\to0$,
since $\lambda\sim M_H^2$ according to Eq.~(\ref{lambda}). 

The tadpole contributions may be quite sizable numerically.
In Ref.~\cite{Jegerlehner:2012kn,Kniehl:2014yia}, it was noted that one-loop
electroweak correction to the $\overline{\mathrm{MS}}$ to pole mass
relationship of the top quark roughly compensates the QCD one.
In particular, the tadpole contribution contains terms that are enhanced
as $N_cM_t^4/(M_W^2 M_H^2)$, where $N_c=3$ is the number of quark colors, at
one loop.
At two loops, such terms appear in square.


\boldmath
\subsection{$\overline{\mathrm{MS}}$ masses}
\label{sec:masses}
\unboldmath

The pole mass $M$ of a particle is defined to be the position of the pole of
its propagator, i.e., the zero of its inverse propagator.\footnote{%
If the particle is unstable, then the pole position has a complex value.
The latter is usually parametrized as
$p^2_\mathrm{pole} = M^2 - i \Gamma M$ for bosons and as
$\sla{p}_\mathrm{pole} = M - i \Gamma/2$ for fermions, where $M$ and $\Gamma$
are the real pole mass and the total decay width of the particle, respectively.
}
For a scalar boson, such as the Higgs boson, one thus needs to solve the
equation
\begin{equation}
\label{invpropH}
  0 = p^2 - m_{H,0}^2 - \Pi_{HH}(p^2) \,,
\end{equation}
where $p$ is the four-momentum, $m_{H,0}$ is the bare mass, and $\Pi_{HH}(p^2)$
is the self-energy function of the Higgs boson.
The solution of Eq.~(\ref{invpropH}) gives the pole mass $M_H$ as a function of
$m_{H,0}$ and other parameters of the SM.
For the $W$ boson, the appropriate equation is similar to Eq.~(\ref{invpropH}),
namely
\begin{equation}
\label{invpropW}
  0 = p^2 - m_{W,0}^2 - \Pi_{WW,T}(p^2) \,,
\end{equation}
where $\Pi_{WW,T}(p^2)$ is the transverse part of the $W$-boson self-energy. 
The $Z$-boson case is somewhat more complicated because of the $\gamma$-$Z$
mixing.
Diagonalizing the corresponding propagator matrix, one obtains
\begin{equation}
\label{invpropZ}
  0 = p^2 - m_{Z,0}^2 - \Pi_{ZZ,T}(p^2) - \frac{\Pi^2_{\gamma Z,T}(p^2)}{p^2 - \Pi_{\gamma\gamma,T}(p^2)}\,,
\end{equation}
where $\Pi_{ZZ,T}(p^2)$, $\Pi_{\gamma Z,T}(p^2)$, and
$\Pi_{\gamma\gamma,T}(p^2)$ are the respective transverse self-energy
functions.
Finally, for a fermion $f$, one has
\begin{eqnarray}
\label{invpropf}
0 =  \sla{p} - m_{f,0} - \Sigma_f(\sla{p}) \,,
\end{eqnarray}
where $\Sigma_f(\sla{p})$ is the self-energy of $f$.\footnote{%
Here, fermion mixing is neglected.
The renormalization for mixed systems of spin-$1/2$ fermions was elaborated in
Ref.~\cite{Kniehl:2012zb}.}
The solutions of Eqs.~(\ref{invpropH})--(\ref{invpropf}) may be found order by
order in perturbation theory, by substituting the ans\"atze
$p^2=m_0^2(1+\kappa_1+\kappa_2+\dots)$ and
$\sla{p}=m_0(1+\kappa_1+\kappa_2+\dots)$, with $\kappa_i$ being the $i$-loop
corrections, in Eqs.~(\ref{invpropH})--(\ref{invpropZ}) and
Eq.~(\ref{invpropf}), respectively.

Alternatively, we may perform the mass renormalization in the
$\overline{\mathrm{MS}}$ scheme.
The relation between the $\overline{\mathrm{MS}}$ mass $m(\mu)$ and $m_0$ has
the simple form
\begin{equation}
   m_0^2 = m^2(\mu) \left(
         1 + \frac{Z^{(1)}}{\varepsilon} + \frac{Z^{(2)}}{\varepsilon^2}
             + \cdots \right)\,,
\label{zm}
\end{equation}
where the expansion parameter $\varepsilon=(4-d)/2$ measures the deviation of
the space-time dimension $d$ from 4.
The mass renormalization constants $Z^{(j)}$ have double expansions in the weak
and strong gauge couplings, $g(\mu)$ and $g_s(\mu)$, respectively.
For the purposes of our two-loop analysis, we need to include the following
terms
\begin{equation}
Z^{(j)}=
\frac{g^2}{16\pi^2}Z^{(j)}_\alpha
+\frac{g^2}{16\pi^2}\,\frac{g_s^2}{16\pi^2}Z^{(j)}_{\alpha\alpha_s}
+\left(\frac{g^2}{16\pi^2}\right)^2Z^{(j)}_{\alpha^2}+\cdots\,.
\end{equation}
As already mentioned in Subsection~\ref{sec:tadpoles}, all the relevant mass
renormalization constants are gauge independent upon the inclusion of the
tadpole contributions.
Explicit expressions through order $O(\alpha^2)$ may be found for the $W$ and
$Z$ bosons in Refs.~\cite{Jegerlehner:2001fb,Jegerlehner:2002em},\footnote{%
The expressions for $Z_{\alpha\alpha_s}^{(1,2)}$ of the $W$ and $Z$ bosons in
Eq.~(4.41) of Ref.~\cite{Jegerlehner:2002em} contain several misprints, which
are corrected in Footnote~9 of Ref.~\cite{Bezrukov:2012sa}.
}
for the top and bottom quarks in Ref.~\cite{Kniehl:2014yia}, and for the Higgs
boson in Appendix~\ref{app:zmmH} of this paper.

In this paper, we shall evaluate the $\overline{\mathrm{MS}}$ masses $m_x(\mu)$
for $x=W,Z,H,t,b$ through order $O(\alpha^2)$.
Besides gauge independence, also the $\mu$ dependence, which is dictated by the
RG, provides a strong check for the correctness of our results.
In fact, the full $\mu$ dependence of $m_x(\mu)$ at two loops may be retrieved from
the one-loop result for $m_x(\mu)$ and its two-loop anomalous dimension.
To simplify the discussion in the remainder of this section, we shall only
allow for one coupling constant, $a(\mu)$.
The generalization to the case under consideration in this paper is
straightforward.

Adopting standard notations, the generic RG equations for $a(\mu)$ and $m(\mu)$
read
\begin{equation}
  \frac{d\,a}{d\ln\mu^2} = \beta \,, \qquad
  \frac{d\, \ln m^2}{d\ln\mu^2} = \gamma \,.
\label{RGeqs}
\end{equation}
The $\beta$ and $\gamma$ functions in Eq.~(\ref{RGeqs}) have perturbative
expansions of the forms $\beta=a^2\beta_1+a^3\beta_2+\cdots$ and
$\gamma=a\gamma_1+a^2\gamma_2+\cdots$.
In the $\overline{\mathrm{MS}}$ scheme, $\beta_j$ are just numbers, while
$\gamma_j$ are, in general, polynomials in squared-mass ratios.
Expressing in turn the $\overline{\mathrm{MS}}$ masses in terms of pole ones,
we find the expression for $\gamma$ to assume the form
\begin{equation}
\gamma = a(\mu) \Gamma_1 + a^2(\mu) ( \Lambda_2 \ln\mu^2 + \Gamma_2 )
+ \cdots \,,
\label{gamma2MM}
\end{equation}
where $\Gamma_i$ and $\Lambda_i$ are independent of $\mu$.
On the other hand, the $\overline{\mathrm{MS}}$ to pole mass relationship has
the form
\begin{equation}
m^2(\mu) = M^2 [ 1 + a(\mu) ( A_1 \ln\mu^2 + B_1 )
         + a^2(\mu) ( A_2 \ln^2\mu^2 + A'_2 \ln\mu + B_2 ) + \cdots ] \,,
\label{mm2MM}
\end{equation}
where $A_i$, $A_i^\prime$, $B_i$, etc., do not depend on $\mu$.
The coefficients $A_1$, $A_2$, and $A'_2$ in front of $\ln^n\mu^2$ with
$n=1,2,\ldots$ may be derived from the RG.
Substituting Eqs.~(\ref{gamma2MM}) and (\ref{mm2MM}) in Eq.~(\ref{RGeqs}) and
comparing the coefficients of $a^m\ln^n\mu^2$, we obtain
\begin{eqnarray}
   A_1 &=& \Gamma_1\,,
\nonumber\\
  2A_2 &=& \Gamma_1^2 + \Lambda_2 - \beta_1\Gamma_1 \,,
\nonumber\\
   A_2^\prime &=& \Gamma_2 + B_1 \Gamma_1 - \beta_1 B_1 \,.
\label{RGresult}
\end{eqnarray}
Consequently, we may independently evaluate the coefficients of $\ln^n\mu^2$
with $n=1,2,\ldots$ from one-loop results and the mass anomalous dimension
$\gamma$.
In the $\overline{\mathrm{MS}}$ scheme, the latter is completely determined by
the single $1/\varepsilon$ pole in the corresponding renormalization constant
in Eq.~(\ref{zm}).
Taking into account both the weak and strong coupling constants, we have
\begin{equation}
  \gamma = \Bigg( \frac{g}{2}\frac{\partial}{\partial g} 
                   + \frac{g_s}{2}\frac{\partial}{\partial g_s} \Bigg) Z^{(1)} \,. 
\end{equation}
On the other hand, $\gamma$ may also be related to the anomalous dimension
$\gamma_{G_F}$ of the Fermi constant $G_F$ and the RG functions in the unbroken
phase of the SM
\cite{Jegerlehner:2001fb,Jegerlehner:2002em,Jegerlehner:2002er}.


\boldmath
\subsection{$\overline{\mathrm{MS}}$ fine-structure constant}
\label{sec:alpha}
\unboldmath

In this section, we discuss different definitions of Sommerfeld's
fine-structure constant $\alpha=e^2/(4\pi)$ and their relationships to the
Fermi constant $G_F$.

The relationship between the $\overline{\mathrm{MS}}$ quantity $\alpha(\mu)$
and the on-shell quantity $\alpha_\mathrm{Th}$ defined in the Thompson limit is
usually written as \cite{Degrassi:1990tu}
\begin{eqnarray}
\alpha(\mu) = \frac{\alpha_{\rm Th}}{1-\Delta\alpha(\mu)} \,, 
\label{alphahat}
\end{eqnarray}
where, to first approximation, $\Delta\alpha(\mu)$ is expressed through the
vacuum polarization function of the photon $\Pi_\gamma^\prime(0)$.
The latter quantity is known to be subject to nonperturbative effects due to
the hadronic content of the photon.
The incorporation of the perturbative $\mathcal{O}(\alpha\alpha_s)$ corrections
was explained in Ref.~\cite{Fanchiotti:1992tu}.
The purely leptonic contributions are known through four loops
\cite{Sturm:2013uka}.
A systematic discussion of higher-order corrections may be found in
Ref.~\cite{Erler:1998sy}.
Through order $\mathcal{O}(\alpha\alpha_s)$, we have\footnote{%
In this work, we actually only need $\alpha(\mu)$ through order $O(\alpha)$.}
\begin{eqnarray}
\Delta\alpha(\mu) &=& \Delta\alpha^{(5)}_{\rm had}(M_Z)
                + \frac{\alpha_{\rm Th}}{4\pi} \left\{
     \left( 7 L_W - \frac{2}{3} \right)
   + \sum_{l=e,\mu,\tau} \left( - \frac{4}{3} L_l \right)\right.
\nonumber\\
&&{}
   + N_c Q_t^2 \left[ - \frac{4}{3} L_t 
          + C_F \frac{\alpha_s}{4\pi} ( - L_t + 15) \right]
\nonumber\\
&&{}
   +\left. N_c \sum_{q=u,d,s,c,b} Q_q^2 \left[ - \frac{4}{3} L_Z + \frac{20}{9}
          + C_F \frac{\alpha_s}{4\pi} \left( - 4 L_Z - 16\zeta_3 
+ \frac{55}{3} \right) \right]
    \right\} + \cdots\,,\qquad
\label{runa}
\end{eqnarray}
where $\Delta\alpha^{(5)}_{\rm had}(M_Z)=0.02772(10)$ \cite{Agashe:2014kda} is the
hadronic contribution to $\Pi_\gamma^\prime(0)$ and $L_x=\ln(M_x^2/\mu^2)$ with
$M_x$ being pole masses.
The four terms inside the curly brackets in Eq.~(\ref{runa}) are the bosonic,
leptonic, top-quark, and perturbative light-quark contributions, respectively.
To the same accuracy as in Eq.~(\ref{runa}), we may also write the $\mu$
dependence of $\alpha(\mu)$ as
\begin{equation}
\alpha(\mu) = \frac{\alpha(M_Z)}{\displaystyle 1 - \frac{\alpha(M_Z)}{4\pi} 
    \left( - \frac{11}{3} - \frac{80}{3}\, \frac{\alpha_s}{4\pi} \right)
\ln\frac{M_Z^2}{\mu^2} + \cdots}\,.
\label{alphahat1}
\end{equation}
Using $\alpha^{-1}(M_Z)=127.940(14)$ \cite{Agashe:2014kda} as input
in Eq.~(\ref{alphahat1}), we obtain $\alpha^{-1}(M_t)=127.565$ at order
$\mathcal{O}(\alpha)$  and $\alpha^{-1}(M_t)=127.540$ at order
$\mathcal{O}(\alpha\alpha_s)$.
The QED $\beta$ function is available through five loops including QCD
corrections \cite{Baikov:2012zm}.

Eqs.~(\ref{alphahat}) and (\ref{runa}) relate $\alpha(M_Z)$, which belongs to
the core of electroweak physics, to the low-energy-QED parameter
$\alpha_{\rm Th}$, which is insensitive to the weak interaction due to the
decoupling of the heavy particles.
While this approach allows one to conveniently implement the RG and so to
consistently accommodate all the QCD contributions of the orders
$O(\alpha\alpha_s^n)$, there are obvious drawbacks if one wishes to go beyond
this approximation.
First, it is less appropriate for the incorporation of electroweak physics,
which comes into play by the exchange of $W$ and $Z$ bosons in the loops.
Second, the definition of $\alpha(\mu)$ from the QED two-point function alone
has only restricted meaning.
To relate $\alpha(\mu)$ to electroweak observables beyond the
one-photon-exchange approximation, one has to include more complicated objects.

A very simple consistent definition of $\alpha(\mu)$ that avoids these
drawbacks is to take Eq.~(\ref{ee}) at face value and to install $G_F$ as an
input parameter via Eqs.~(\ref{g2}) and (\ref{g2g2}), viz.\
\begin{eqnarray}
\alpha(\mu) = \frac{\sqrt{2}G_F M_W^2}{\pi}
   [ 1 + \delta_W(\mu) ]
   \left[ 1 - \frac{M_W^2}{M_Z^2}\, \frac{1 + \delta_W(\mu)}{1 + \delta_Z(\mu)}
 \right] \,,
\label{alphamu}
\end{eqnarray}
where $\delta_W(\mu)$ and $\delta_Z(\mu)$ are given in Eqs.~(\ref{g2}) and
(\ref{g2g2}), respectively.
This definition of $\alpha(\mu)$ is exact and gauge independent.
Since $\delta_W(\mu)$ and $\delta_Z(\mu)$ themselves depend on $\alpha(\mu)$,
Eq.~(\ref{alphamu}) provides an implicit definition of $\alpha(\mu)$, which
may be solved iteratively, e.g., using the Newton-Raphson method.
Working through two loops, we thus obtain numerically at $\mu=M_Z$
  \begin{eqnarray}  
    \alpha^{-1}(M_Z) = 132.233 - \underbrace{4.741}_{\alpha} 
    + \underbrace{0.512}_{\alpha\alpha_s} 
    + \underbrace{0.203}_{\alpha^2} = 128.208 \,, 
    \label{a1}
  \end{eqnarray}
where the first number on the right-hand side is the tree-level value
$\alpha_F=\sqrt{2}G_F M_W^2(1-M_W^2/M_Z^2)/\pi=1/132.233\dots$ and the
$\mathcal{O}(\alpha)$, $\mathcal{O}(\alpha\alpha_s)$, and
$\mathcal{O}(\alpha^2)$ contributions are specified separately.
Alternatively we can expand Eq.~(\ref{alphamu}) as a power series and truncate 
it beyond $O(\alpha^2)$, as
\begin{eqnarray}
\alpha(\mu) = \alpha_F (1 + \alpha_F X_{10}(\mu) + \alpha_F\alpha_s(\mu) X_{11}(\mu) 
           + \alpha_F^2 X_{20}(\mu) + \dots) \,,
\end{eqnarray}
with analytic coefficients $X_{ij}(\mu)$.
Numerically, this gives
  \begin{eqnarray}
    \alpha^{-1}(M_Z) = 132.233 - \underbrace{4.648}_{\alpha} 
    + \underbrace{0.491}_{\alpha\alpha_s} 
    + \underbrace{0.090}_{\alpha^2} = 128.166 \,.
    \label{a2}
  \end{eqnarray}
The difference between the values in Eqs.~(\ref{a1}) and (\ref{a2}) may be
interpreted as the theoretical uncertainty due to higher-order effects.


\section{Results and discussion}
\label{sec:results}

We are now in a position to present our numerical analysis.
For this, we adopt the following values of the input parameters from the
Particle Data Group \cite{Agashe:2014kda}:
\begin{eqnarray}
  \label{eq:SMinput}
  & M_W = 80.385(15)\, \mbox{GeV}, \quad 
        M_Z = 91.1876(21) \, \mbox{GeV}, \quad 
        M_H = 125.7(4) \, \mbox{GeV}, \nonumber \\
  & M_t = 173.21(51)(71) \, \mbox{GeV}, \quad 
       M_b = 4.9 \, \mbox{GeV}, \nonumber\\ 
  &   G_F = 1.166 378 7(6)\times  10^{-5} \, \mbox{GeV}^{-2}, \nonumber \\
  & \alpha^{-1}(M_Z) =  127.940(14), \quad \alpha_s(M_Z) = 0.1185(6) \,.
\end{eqnarray}
We neglect the masses $M_f$ of the light fermions $f\neq t,b$, since their effects
are negligible and do not play any r\^ole in our considerations.
The mass of the bottom quark is taken into account in all one-loop expressions,
but is neglected in higher-order terms.
The order $\mathcal{O}(\alpha\alpha_s M_b^2/M_W^2)$ of the discarded terms is
far beyond the accuracy of the two-loop approximations.
In most cases, we consider the threshold corrections in
Eqs.~(\ref{g2})--(\ref{y2}) at a relatively high matching scale, of the order
of $M_Z$ or $M_t$.
However, in the case of the bottom quark, it is more natural to choose a
somewhat lower matching scale, of the order of $M_b$.
Therefore, we consider $y_b$ and $m_b$ separately.
Evolving the value of $\alpha(M_Z)$ in Eq.~(\ref{eq:SMinput}) with the help of
Eq.~(\ref{runa}) at order $O(\alpha\alpha_s^4)$ to the scale $\mu=M_t$, we
obtain $\alpha(M_t)=127.540$, as already quoted in Section~\ref{sec:alpha}.
Applying four-loop evolution and three-loop matching
\cite{Schroder:2005hy} to the value of $\alpha_s^{(5)}(M_Z)$ in
Eq.~(\ref{eq:SMinput}), we obtain
$\alpha_s^{(6)}(M_t) = 0.1081$.

\subsection{Threshold corrections to the couplings and masses}
\label{sec:interp-yukawa}

We first present the threshold corrections $\delta_x$ with $x=W,Z,H,t$,
which we parametrize as follows
\begin{equation}
  1 + \delta_x(\mu)  = 1 + \frac{\alpha(\mu)}{4\pi} Y_x^{1,0} 
    + \frac{\alpha(\mu)}{4\pi}\, \frac{\alpha_s(\mu)}{4\pi} Y_x^{1,1} 
    + \left(\frac{\alpha(\mu)}{4\pi} \right)^2 Y_x^{2,0} + \cdots\,.
\label{eq:2}
\end{equation}
The coefficients $Y_x^{1,0}$, $Y_x^{1,1}$, and $Y_x^{2,0}$
may be evaluated numerically using a computer program to be published in
a forthcoming paper \cite{cpc}.
In Appendix~\ref{app:res}, we list all the $\mathcal{O}(\alpha)$ and
$\mathcal{O}(\alpha\alpha_s)$ coefficients, $Y_x^{1,0}$ and $Y_x^{1,1}$, in
their full analytic forms and the $\mathcal{O}(\alpha^2)$ coefficients,
$Y_x^{2,0}$, in the gaugeless-limit approximation.
For the reader's convenience, we present linear interpolation formulae for the
two-loop coefficients, $Y_x^{1,1}$ and $Y_x^{2,0}$, at the two most important
matching scales, $\mu=M_Z,M_t$.
It is understood that the variables $M_t$ and $M_H$ are to be taken in units of
GeV.
At $\mu=M_Z$, we have
\begin{align}
  \label{eq:interplationsMZ}
  %
  %
  %
  %
   Y_W^{1,1} & = &
  -0.277897 & \;(M_t - 173.2) & {}+ 0 & \;(M_H - 125.7) &
  {}+27.8149\,, \nonumber \\
   Y_W^{2,0} & = &
  4.83906 & \;(M_t - 173.2) & {}+ 1.66627 & \;(M_H - 125.7) &
  {}+ 2740.81\,, \nonumber \\
  %
  %
  %
  %
   Y_Z^{1,1} & = &
  -2.15581 & \;(M_t - 173.2) & {}+ 0 & \;(M_H - 125.7) &
  {}-180.9\,, \nonumber \\
   Y_Z^{2,0} & = &
  -35.6751 & \;(M_t - 173.2) & {}+ 1.12839 & \;(M_H - 125.7) &
  {}+ 729.793\,, \nonumber \\
  %
  %
  %
  %
   Y_H^{1,1} & = &
  -85.325 & \;(M_t - 173.2) & {}+ 67.8568 & \;(M_H - 125.7) &
  {}-3770.21\,, \nonumber \\
   Y_H^{2,0} & = &
   -2487.1 & \;(M_t - 173.2) & {}+ 917.155 & \;(M_H - 125.7) &
  {}-54051.1\,, \nonumber \\
  %
  %
  %
  %
   Y_t^{1,1} & = &
  -5.09107 & \;(M_t - 173.2) & {}+ 0.800608 & \;(M_H - 125.7) &
  {}-166.005\,, \nonumber \\
   Y_t^{2,0} & = &
  98.6532 & \;(M_t - 173.2) & {}+ 5.47244 & \;(M_H - 125.7) &
  {}+ 3318.95  \,.
\end{align}
At $\mu=M_t$, we have
\begin{align}
  \label{eq:interplationsMt}
  %
  %
  %
  %
   Y_W^{1,1} & = &
  -0.277897 & \;(M_t - 173.2) & {}+ 0 & \;(M_H - 125.7) &
  {}+96.8963\,, \nonumber \\
   Y_W^{2,0} & = &
  18.8904 & \;(M_t - 173.2) & {}+ 1.2212 & \;(M_H - 125.7) &
  {}+ 4714.29\,, \nonumber \\
  %
  %
  %
  %
   Y_Z^{1,1} & = &
  -2.15581 & \;(M_t - 173.2) & {}+ 0 & \;(M_H - 125.7) &
  {}-121.818\,, \nonumber \\
   Y_Z^{2,0} & = &
  -22.9901 & \;(M_t - 173.2) & {}+ 0.67285 & \;(M_H - 125.7) &
  {}+ 2450.17\,, \nonumber \\ 
  %
  %
  %
  %
   Y_H^{1,1} & = &
  -93.1689 & \;(M_t - 173.2) & {}+ 40.2044 & \;(M_H - 125.7) &
  {}-1933.13\,, \nonumber \\
   Y_H^{2,0} & = &
  -407.022 & \;(M_t - 173.2) & {}+ 28.456 & \;(M_H - 125.7) &
  {}-6753.38\,, \nonumber \\
  %
  %
  %
  %
   Y_t^{1,1} & = &
  -0.304831 & \;(M_t - 173.2) & {}+ 1.02857 & \;(M_H - 125.7) &
  {}-79.4078\,, \nonumber \\
   Y_t^{2,0} & = &
  19.6085 & \;(M_t - 173.2) & {}+ 1.64258 & \;(M_H - 125.7) &
  {}+ 699.287 \,.
\end{align}

We now present the $\overline{\mathrm{MS}}$ to pole mass relationships, in
forms similar to Eq.~(\ref{eq:2}).
Specifically, we write
\begin{equation}
  \frac{m_B^2(\mu)}{M_B^2} = 1 + \frac{\alpha(\mu)}{4\pi} X_B^{1,0} 
  + \frac{\alpha(\mu)}{4\pi}\, \frac{\alpha_s(\mu)}{4\pi} X_B^{1,1} 
  + \left(\frac{\alpha(\mu)}{4\pi}\right)^2 X_B^{2,0} + \cdots  
\label{eq:massbos}
\end{equation}
for the bosons $B=W,Z,H$ and
\begin{eqnarray}
  \frac{m_f(\mu)}{M_f} &=& 1 + \frac{\alpha(\mu)}{4\pi} X_f^{1,0} 
  + \frac{\alpha(\mu)}{4\pi}\, \frac{\alpha_s(\mu)}{4\pi} X_f^{1,1} 
  + \left(\frac{\alpha(\mu)}{4\pi}\right)^2 X_f^{2,0} + \dots  
\label{eq:massfer} 
\end{eqnarray}
for the fermion $f=t$.
The coefficients $X_x^{1,0}$, $X_x^{1,1}$, and $X_x^{2,0}$
may be evaluated numerically using the computer
program to be released \cite{cpc}.
Convenient interpolation formulae for the two-loop coefficients, $X_x^{1,1}$ and
$X_x^{2,0}$, at $\mu=M_Z$ read
\begin{align}
  \label{eq:massintMZ}
   X_W^{1,1} & = &
  43.129 & \;(M_t - 173.2) & {}-37.6735 & \;(M_H - 125.7) & 
  {}+ 2287.33\,, \nonumber \\
  X_W^{2,0} & = &
  2409.02 & \;(M_t - 173.2) & {}-727.366 & \;(M_H - 125.7) & 
  {}+ 33860.5\,, \nonumber \\
   X_Z^{1,1} & = &
  41.2511 & \;(M_t - 173.2) & {}-37.6735 & \;(M_H - 125.7) & 
  {}+ 2078.62\,, \nonumber \\
  X_Z^{2,0} & = &
  2285.27 & \;(M_t - 173.2) & {}-704.11 & \;(M_H - 125.7) & 
  {}+ 31005.9\,, \nonumber \\
   X_H^{1,1} & = &
  -41.9182 & \;(M_t - 173.2) & {}+ 30.1834 & \;(M_H - 125.7) & 
  {}-1510.7\,, \nonumber \\
  X_H^{2,0} & = &
  -1214.38 & \;(M_t - 173.2) & {}+ 636.554 & \;(M_H - 125.7) & 
  {}-31764.7\,, \nonumber \\
   X_t^{1,1} & = &
  15.9663 & \;(M_t - 173.2) & {}-18.1532 & \;(M_H - 125.7) & 
  {}+ 967.983\,, \nonumber \\
  X_t^{2,0} & = &
  1288.64 & \;(M_t - 173.2) & {}-352.559 & \;(M_H - 125.7) & 
  {}+ 18890.7 \,. 
\end{align}

Comparing Eqs.~(\ref{eq:interplationsMZ}) and (\ref{eq:massintMZ}), we observe
that the coefficients $X_{W,Z,t}^{1,1}$ and $X_{W,Z,t}^{2,0}$ are much larger
than their counterparts $Y_{W,Z,t}^{1,1}$ and $Y_{W,Z,t}^{2,0}$, typically by
an order of magnitude.
This is due to the tadpole contributions that dominate the former coefficients,
but largely cancel in the latter, as already discussed in
Subsection~\ref{sec:tadpoles}.
In the case of the Higgs boson, however, the situation is quite different.
In fact, $X_H^{(1,1)}$ and $X_H^{(2,0)}$ are in the same ballpark as
$Y_H^{(1,0)}$ and $Y_H^{(2,0)}$, respectively, because the tadpole cancellation
does not take place.

\begin{table}[h!]
\centering
\caption{QCD, $\mathcal{O}(\alpha)$, $\mathcal{O}(\alpha\alpha_s)$, and
$\mathcal{O}(\alpha^2)$ contributions to $m_t(M_t)-M_t$ in GeV for
$M_H=124,125,126$~GeV.
The QCD contribution includes the orders $\mathcal{O}(\alpha^n)$ with
$n=1,2,3$.
The numbers in parentheses are obtained in the gaugeless-limit approximation.}
\label{tab:mt-Mt}
\medskip
\begin{tabular}{|c|c|c|c|c|c|}
    \hline
    $M_H$ [GeV] & QCD & $\mathcal{O}(\alpha)$ & $\mathcal{O}(\alpha\alpha_s)$
 & $\mathcal{O}(\alpha^2)$ & total  \\ \hline
    124 & $-10.38$ & 12.08 & $-0.39$ & $-0.99$ ($-0.47$) & 0.32 \\ \hline
    125 & $-10.38$ & 11.88 & $-0.39$ & $-0.96$ ($-0.45$) & 0.14 \\ \hline
    126 & $-10.38$ & 11.67 & $-0.38$ & $-0.94$ ($-0.44$) & $-0.03$ \\ \hline
  \end{tabular}
\end{table}

Let us now consider the various contributions to the $m_t(M_t)-M_t$ shift.
In Ref.~\cite{Kniehl:2014yia}, we presented a detailed analysis thereof,
providing the contribution of order $\mathcal{O}(\alpha^2)$ in the
gaugeless-limit approximation and comparing it with the well-known
contributions of orders $\mathcal{O}(\alpha_s^n)$ with $n=1,2,3$,
$\mathcal{O}(\alpha)$, and $\mathcal{O}(\alpha\alpha_s)$.
Our results were summarized in Table~1 of Ref.~\cite{Kniehl:2014yia}.
Here, we improve this analysis by including the full $\mathcal{O}(\alpha^2)$
contribution and updating the input parameters as specified in
Eq.~(\ref{eq:SMinput}).
Our new results are presented in Table~\ref{tab:mt-Mt}.
We observe that the gaugeless-limit approximation successfully predicts the
true sign of the $\mathcal{O}(\alpha^2)$ contribution, but significantly
underestimates its magnitude by accounting for only about one half of it.
The seemingly poor quality of this approximation may be partly ascribed to the
presence of a heavy particle on the external lines of the two-point diagrams
involved.
On the other hand, the absolute deviation by itself may be considered 
acceptable.

\begin{table}[h!]
\centering
\caption{Same as in Table~\ref{tab:mt-Mt}, but for $\delta_t(M_t)$ in units of
$10^{-4}$.}
\label{tab:yt}
\medskip
  \begin{tabular}{|c|c|c|c|c|c|c|}
    \hline
    $M_H$ [GeV] & QCD & $\mathcal{O}(\alpha)$ & $\mathcal{O}(\alpha\alpha_s)$ &
    $\mathcal{O}(\alpha^2)$ & total \\ \hline
    124 & $-599.3$ & 13.5 & $-4.4$ & 2.7 (3.1) & $-587.4$ \\ \hline
    125 & $-599.3$ & 13.2 & $-4.3$ & 2.7 (3.1) & $-587.7$ \\ \hline
    126 & $-599.3$ & 12.9 & $-4.2$ & 2.7 (3.1) & $-587.9$ \\ \hline
  \end{tabular}
\end{table}

In Table~\ref{tab:yt}, the analysis of Table~\ref{tab:mt-Mt} is repeated for
$\delta_t(M_t)$.
As expected by reason of the tadpole cancellation, the QCD corrections are by
far dominant.
In fact, the electroweak corrections only make up approximately 2\% of
$\delta_t(M_t)$.
Furthermore, we observe that the electroweak perturbative expansion exhibits a
useful convergence behavior.
In contrast to the case of the $m_t(M_t)-M_t$ shift, the gaugeless-limit
approximation works well, overestimating the true $\mathcal{O}(\alpha^2)$
correction by about 15\%.

\begin{table}[h!]
\centering
\caption{Same as in Table~\ref{tab:mt-Mt}, but for $\delta_H(M_t)$ in units of
$10^{-4}$.
There are no pure QCD corrections in this case.}
\label{tab:lam}
\medskip
  \begin{tabular}{|c|c|c|c|c|c|}
    \hline
    $M_H$ [GeV] & $\mathcal{O}(\alpha)$ & $\mathcal{O}(\alpha\alpha_s)$ &
    $\mathcal{O}(\alpha^2)$ & total \\ \hline
    124 & $-114.8$ & $-107.5$ & $-26.6$ ($-29.1$) & $-248.7$ \\ \hline
    125 & $-114.5$ & $-105.2$ & $-26.4$ ($-29.2$) & $-246.1$ \\ \hline
    126 & $-114.1$ & $-103.1$ & $-26.3$ ($-29.3$) & $-243.5$ \\ \hline
  \end{tabular}
\end{table}

In Table~\ref{tab:lam}, the corresponding results are presented for
$\lambda(M_t)$,
Of course, there are no pure QCD contributions in this case.
As already observed in Ref.~\cite{Bezrukov:2012sa}, the
$\mathcal{O}(\alpha\alpha_s)$ contribution almost doubles the
$\mathcal{O}(\alpha)$ one.
The $\mathcal{O}(\alpha^2)$ contribution reaches about one quarter of the
$\mathcal{O}(\alpha)$ one.
The gaugeless-limit approximation works here even better than in the case of
$\delta_t(M_t)$, with a deviation of about 10\%.

Finally, we study the quantities related to the bottom quark, at $\mu=M_b$.
For the $\overline{\mathrm{MS}}$ to pole mass relationship, we obtain
\begin{equation}
\{m_b(M_b) - M_b \}_{{\rm QCD},O(\alpha),O(\alpha\alpha_s),O(\alpha^2)} = 
         -0.85 - 1.90  - 1.53 + 1.75\ (1.80) \mbox{GeV} \,.
\label{eq:mbshift}
\end{equation}
In Eq.~(\ref{eq:mbshift}), the electroweak corrections are overwhelming and
do not exhibit a useful convergence behavior, again because of the uncanceled
tadpole contributions.
The latter render the electroweak extension of the $\overline{\mathrm{MS}}$
mass definition for the bottom quark quite unfeasible in practice, while they
do not affect its pole mass.
This situation is unfamiliar from pure QCD, which is tadpole free.
Here, the $\overline{\mathrm{MS}}$ mass is frequently preferred to the
pole mass because it is only sensitive to short-distance effects, while the
latter suffers from a renormalon ambiguity.
This is reflected by the relatively large size of the first term on the
right-hand side of Eq.~(\ref{eq:mbshift}) and the slow convergence of the
$\mathcal{O}(\alpha_s^n)$ corrections with $n=1,2,3$ which build it up.
We also learn from Eq.~(\ref{eq:mbshift}) that the gaugeless-limit
approximation works here remarkably well, within an error of about 3\%.
On the other hand, the threshold corrections to the Yukawa coupling,
\begin{equation}
\{1+\delta_b(M_b)\}_{{\rm QCD},O(\alpha),O(\alpha\alpha_s),O(\alpha^2)} = 
      1 - 0.1728 - 0.0190 - 0.0112 + 0.0032(0.0033) \,,
\end{equation}
are perturbatively stable as for the tadpole contributions. 
Also here, the gaugeless-limit approximation works at the 3\% level.


\section{Conclusions}

In this paper, we analytically evaluated, through two loops in the SM, the
threshold corrections to the electroweak gauge couplings, the top and bottom
Yukawa couplings, the quartic self-coupling, and the vacuum expectation value
of the scalar field as well as the $\overline{\mathrm{MS}}$ to pole mass
shifts of the top and bottom quarks.
Specifically, we included the corrections of orders
$\mathcal{O}(\alpha)$, $\mathcal{O}(\alpha\alpha_s)$, and
$\mathcal{O}(\alpha^2)$.
We emphasized the importance of the tadpole contributions to render these
corrections gauge independent.
Besides comparisons with the literature, UV finiteness and RG invariance served
as further checks of the correctness of our results.

The threshold corrections to the gauge and Yukawa couplings are finite in the
limit of vanishing Higgs-boson mass $M_H$ due to cancellations of leading
tadople contributions, and their perturbative expansions exhibit useful
convergence.
The threshold correction $\delta_H(\mu)$ to the quartic scalar self-coupling
$\lambda(\mu)$ scales as $1/M_H^2$ for $M_H\to0$, but $\lambda(\mu)$ is finite
in this limit.
Also in this case, perturbative stability is intact.
All these threshold corrections are central ingredients for RG analyses within
the SM and, in particular, for the determination of the $M_H$ lower bound from
the requirement of vacuum stability at the scale of the Planck mass
\cite{Bezrukov:2012sa,Degrassi:2012ry,Buttazzo:2013uya}.
By contrast, the $\overline{\mathrm{MS}}$ to pole mass shifts, which do not
enter such RG analyses, suffer from sizable tadpole contributions, which are
particularly severe in the case of the bottom quark.
As a way out of this problem, it was proposed in
Refs.~\cite{Jegerlehner:2012kn,Kniehl:2014yia} to define the running fermion
masses in terms of the $\overline{\mathrm{MS}}$ Yukawa couplings, as
$m_{Y,f}(\mu)=2^{-3/4}G_F^{-1/2}y_f(\mu)$.

\section*{Acknowledgments}

We thank A. Bednyakov for fruitful discussions and F. Jegerlehner for a
useful communication regarding $\Delta\alpha(\mu)$.
This work was supported in part by the German Federal Ministry for Education
and Research BMBF through Grant No.\ 05H12GUE and by the German Research
Foundation DFG through the Collaborative Research Centre No.\ 676
{\it Particles, Strings and the Early Universe: the Structure of Matter and
Space-Time}. The work of AP is supported by Russian President
Grant No. MK-1001.2014., JINR Grant for young scientists No. 15-302-07
and the Dynasty Foundation.

\appendix


\section{Analytical results}
\label{app:res}

In this appendix, we present analytical results for the threshold corrections
$\Delta\overline{r}$ and $\delta_x$ with $x=W,Z,H,t,b$ defined in
Eqs.~(\ref{dRdef0})--(\ref{y2}).
The $\overline{\mathrm{MS}}$ to pole mass relationships for these bosons and
fermions then follow according to Eq.~(\ref{mfer}).

Let us first introduce the notations.
We work in dimensional regularization with $d=4-2\varepsilon$ space-time
dimensions and 't~Hooft mass $\mu$.
$N_c=3$ is the number of quark colors, $C_F = (N_c^2-1)/(2N_c)=4/3$, and
$n_G=3$ is the number of fermion generations.
We express our results in terms of the pole masses and use the abbreviations
\begin{eqnarray}
  w &=& M_W^2 ,\qquad
  z = M_Z^2 ,\qquad
  h = M_H^2 ,\qquad
  t = M_t^2 ,\qquad
  b = M_b^2 ,\qquad
\nonumber\\
S_w^2 &=& 1 - w/z,\qquad
L_x = \ln(x/\mu^2)\qquad (x=w,z,h,t,b).
\end{eqnarray}

As is well known, any one-loop two-point Feynman integral can be expressed in
terms of two types of master integrals, namely
\begin{eqnarray}
   {\bf A}(u_1) &=& \int \frac{d\tilde{k}}{k^2-u_1} \,,
\nonumber
\\
   {\bf B}(q^2;u_1,u_2) &=& \int \frac{d\tilde{k}}{[k^2-u_1][(k-q)^2-u_2]} \,,
\label{Bbold}
\end{eqnarray}
where
\begin{eqnarray}
   d\tilde{k} = (e^{-\gamma_E}\mu^2)^{\varepsilon} \frac{d^dk}{\pi^{d/2}}  
\end{eqnarray}
is the integral measure in $d$-dimensional Minkowski space-time, $u_j$ are the
squares of the masses of the internal propagators, and $q$ is the external
four-momentum.
To suppress the appearance of Euler's constant $\gamma_E$ in the Laurent
expansions in $\varepsilon$, we pull out the factor
$e^{-\gamma_E\varepsilon}$ for each loop momentum.

In the following, we shall need the expansions of the Feynman integrals in
Eq.~(\ref{Bbold}) in $\varepsilon$ through order $O(\varepsilon)$.
They read
\begin{eqnarray}
{\bf A}(u_1) &=& i \left[ \frac{u_1}{\varepsilon} - A_0(u_1) - \varepsilon A_{0,\varepsilon}(u_1) 
             + O(\varepsilon^2) \right]\,,
\nonumber\\
{\bf B}(q^2; u_1, u_2) &=& i \left[ \frac{1}{\varepsilon} + B_0(q^2;u_1,u_2) 
             + \varepsilon B_{0,\varepsilon}(q^2;u_1,u_2) 
             + O(\varepsilon^2) \right]\,,
\label{eq:ab}
\end{eqnarray}
where
\begin{eqnarray}
  A_0(u_1) &=& u_1 \left( \ln\frac{u_1}{\mu^2} - 1 \right) \,,
\\
  A_{0,\varepsilon}(u_1) &=& u_1 \left( - 1 - \frac{1}{2}\zeta_2 +
 \ln\frac{u_1}{\mu^2} - \frac{1}{2} \ln^2\frac{u_1}{\mu^2} \right) \,,
\\
\label{B0def}
  B_0(s; u_1, u_2) &=& - \int\limits_0^1 dx\,
     \ln \frac{x u_1 + (1-x) u_2 - x(1-x)s}{\mu^2} \,,
\\
  B_{0,\varepsilon}(s; u_1, u_2) &=& \frac{1}{2}\zeta_2 + \frac{1}{2} \int\limits_0^1 dx
     \ln^2\frac{x u_1 + (1-x) u_2 - x(1-x)s}{\mu^2} \,.
\end{eqnarray}
The sign conventions in Eq.~(\ref{eq:ab}) have been chosen in accordance with
the program library {\tt TSIL} \cite{Martin:2005qm}.
Some care must be exercised when a complex-valued function $B_0$ gets squared
or multiplied by some other complex-valued $B_0$ function.
In particular, one needs to distinguish $[\re B_0(\dots)]^2$ from
$\re B_0^2(\dots)$.
In such cases, we explicitly indicate when the real part should be taken.
In particular, we have
\begin{equation}
  B_0( t; 0,0) = \re B_0( t; 0,0) - i\pi ,
\end{equation}
if $t>0$, and 
\begin{equation}
  B_0( t; 0,w) = \re B_0( t; 0,w) - i\pi \frac{t-w}{t},
\end{equation}
if $t>w>0$. 
The former situation occurs, e.g., for $\delta_H$ and $\delta_Z$ and the latter
for $\delta_t$.

At two loops, the most general form of the self-energy integral is given by
\begin{eqnarray}
\lefteqn{J_{a_1 a_2 a_3 a_4 a_5}(q^2;u_1,u_2,u_3,u_4,u_5)=}
\nonumber\\
&&{}  \int
     \frac{d\tilde{k}_1\, d\tilde{k}_2}{[k_1^2-u_1]^{a_1}[k_2^2-u_2]^{a_2}
        [(k_1-q)^2-u_3]^{a_3}[(k_2-q)^2-u_4]^{a_4}[(k_1-k_2)^2-u_5]^{a_5}} \,,
\label{Jdef}
\end{eqnarray}
where $a_j$ indicate the powers of the respective propagators.
The numerator of a Feynman integral may be incorporated in the representation
of Eq.~(\ref{Jdef}) by allowing for some of the indices $a_j$ to be negative.
Alternatively, one may reduce any tensor integral to scalar ones in higher
(shifted) space-time dimensions \cite{Tarasov:1996br}.
Then, the recurrence relations of Ref.~\cite{Tarasov:1997kx} may be used to
shift the indices as well as the space-time dimension to the appropriate
values.
After the reduction, any two-loop two-point Feynman integral may be expressed
as a linear combination of a finite set of master integrals with the
coefficients being rational functions of $q^2$, $u_j$, and $d$. 
However, the choice of the master integrals is not unique.
In particular, our set of master integrals differs from the one in
Ref.~\cite{Tarasov:1997kx}.
A drawback of the set in Ref.~\cite{Tarasov:1997kx} is that the coefficients in
front of master integrals may contain poles in $\varepsilon$. 
When this happens, then deeper $\varepsilon$ expansions of the master integrals
are required.\footnote{%
See, e.g., the explicit calculation in Ref.~\cite{Fleischer:1998dw},
             based on the setup of Ref.~\cite{Tarasov:1997kx},
             where poles in $\varepsilon$ through the second order arose
             and some the two-loop master integrals needed to be expanded
             through $O(\varepsilon^2)$.}
Furthermore, the master integrals themselves may posses infrared divergences
besides the ultraviolet ones.
It is possible, however, to choose the master integrals in such a way that: 
(a) the coefficients in front of the master integrals always have smooth limits
$\epsilon\to0$ and (b) the master integrals themselves are infrared finite.
Thanks to property (a), we do not need to expand master integrals beyond the
order $O(\varepsilon^0)$.

We keep our notations for the master integrals very close to those of
Ref.~\cite{Martin:2005qm}.
Specifically, the different integrals in Eq.~(\ref{Jdef}) are denoted by the
bold-faced letters $\mathbf{I}$, $\mathbf{S}$, $\mathbf{T}$, $\mathbf{U}$,
$\mathbf{V}$, and $\mathbf{M}$.
They always have indices $a_j=0,1$, except for $\mathbf{T}$ and $\mathbf{V}$,
which have $a_j=2$ one time.
These functions generally have ultraviolet poles in $\varepsilon$.
We denote the $O(\varepsilon^0)$ terms of these master integrals by
$I_0$, $S_0$, $T_0$, $U_0$, $V_0$, and $M_0$, respectively.\footnote{%
The difference between these functions and their counterparts
$I$, $S$, $T$, $U$, $V$, and $M$ defined in Ref.~\cite{Martin:2005qm} may be
gleaned from Eqs.~(2.34)--(2.39) therein.}
The arguments of all these functions are written as in
Ref.~\cite{Martin:2005qm}, except that we always include $q^2$ as the first
argument if it is present.
As in Ref.~\cite{Martin:2005qm}, we exclude $\mu^2$ from the argument lists.

Specifically, $I_0$, $S_0$, $T_0$, $U_0$, $V_0$, and $M_0$ are defined via
Eq.~(\ref{Jdef}) as
\begin{eqnarray}
&&{}
J_{11111}(q^2;u_1,u_2,u_3,u_4,u_5) = M_0(q^2;u_1,u_2,u_3,u_4,u_5) + O(\varepsilon)\,,
\nonumber\\
&&{}
J_{01101}(q^2;u_1,u_2,u_3,u_4,u_5) = - \frac{u_2+u_3+u_5}{2\varepsilon^2}
\nonumber\\
&&\qquad{}
      + \left( A_0(u_2)+A_0(u_3)+A_0(u_5)- \frac{u_2+u_3+u_5}{2} + \frac{q^2}{4} \right) \frac{1}{\varepsilon}
\nonumber\\
&&\qquad{}
      + S_0(q^2;u_2,u_3,u_5) + O(\varepsilon)\,,
\nonumber\\
&&{}
J_{02101}(q^2;u_1,u_2,u_3,u_4,u_5) = - \frac{1}{2\varepsilon^2}
      + \left( \frac{A_0(u_2)}{u_2} + \frac{1}{2} \right) \frac{1}{\varepsilon}
\nonumber\\
&&\qquad{}
      - T_0(q^2;u_2,u_3,u_5) + O(\varepsilon)\,,
\nonumber\\
&&{}
J_{11001}(q^2;u_1,u_2,u_3,u_4,u_5) = - \frac{u_1+u_2+u_5}{2\varepsilon^2}
\nonumber\\
&&\qquad{}
\left(A_0(u_1)+A_0(u_2)+A_0(u_5)-\frac{u_1+u_2+u_5}{2}\right)
\frac{1}{\varepsilon}
      + I_0(u_1,u_2,u_5) + O(\varepsilon)\,,
\nonumber\\
&&{}
J_{11101}(q^2;u_1,u_2,u_3,u_4,u_5) = - \frac{u_1+u_2+u_5}{2\varepsilon^2}
\nonumber\\
&&\qquad{}
      - \frac{1}{2\varepsilon^2}
      - \left( \frac{1}{2} + B_0(q^2;u_1,u_3) \right) \frac{1}{\varepsilon}
      - U_0(q^2;u_3,u_1,u_5,u_2) + O(\varepsilon)\,,
\nonumber\\
&&{}
J_{21101}(q^2;u_1,u_2,u_3,u_4,u_5) = \Bigg(
        (q^2+u_1-u_3) \Big( B_0(q^2;u_1,u_3)-1 \Big) 
\nonumber\\
&&\qquad{}
           + 2 A_0(u_1)
              + \frac{q^2-u_1-u_3}{u_3}A_0(u_3) \Bigg)
          \frac{1}{\Delta(q^2,u_1,u_3)} \frac{1}{\varepsilon}
\nonumber\\
&&\qquad{}
      + V_0(q^2;u_3,u_1,u_5,u_2) + O(\varepsilon)\,,
\end{eqnarray}
where $\Delta(x,y,z)=x^2+y^2+z^2-2xy-2yz-2zx$.
As in the one-loop case, the above master integrals may develop imaginary
parts, while our threshold corrections take strictly real values. 
Since the two-loop functions never appear in products with complex
coefficients, we impose the rule that their imaginary parts are to be
discarded.

In Ref.~\cite{Kniehl:2014yia}, the threshold corrections $\Delta\overline{r}$,
$\delta_t$, and $\delta_b$ were calculated at $\mathcal{O}(\alpha^2)$ in the
gaugeless-limit approximation and expressed in terms of the functions
${\cal H}(x)$ and $\Phi(x)$ defined, respectively, in Eqs.~(41) and (42)
therein.
The latter are related to the function $I_0$ introduced above with specific
arguments, namely
\begin{eqnarray}
I_0(h,t,t) &=&
    \frac{1}{2}(4t-h)\, \Phi\left( \frac{h}{4t} \right)
    - \frac{1}{2}(7 + \zeta_2 )( 2t+h)
    + 6 t \ln\frac{t}{\mu^2}
    + 3 h \ln\frac{h}{\mu^2}
\nonumber\\
&&{}
    - h \ln\frac{h}{\mu^2} \ln\frac{t}{\mu^2}
    - \frac{1}{2} h \ln^2\frac{h}{\mu^2}
    - \frac{1}{2}(4t-h) \ln^2\frac{t}{\mu^2}  \,,
\nonumber\\
I_0(0,h,t) &=&
    - \frac{t-h}{2} \, {\cal H}\left( \frac{h}{t} \right)
    - \frac{1}{2}(7 + \zeta_2 )(t+h)
    + 3 t \ln\frac{t}{\mu^2}
    + 3 h \ln\frac{h}{\mu^2}
\nonumber\\
&&{}
    + \frac{t+h}{4} \left( \ln\frac{h}{\mu^2} + \ln\frac{t}{\mu^2} \right)^2
    - h \ln^2\frac{h}{\mu^2}
    - t \ln^2\frac{t}{\mu^2}  \,.
\end{eqnarray}
Similarly, the functions $I_1$ and $I_2$ defined in Eq.~(18) of
Ref.~\cite{Kniehl:2014yia} are related to the function $B_0$ defined in
Eq.~(\ref{B0def}) above, as
\begin{eqnarray}
B_0(t;h,t) &=& - \ln\frac{t}{\mu^2} - I_1\left(\frac{h}{t}\right) \,,
\nonumber \\
B_0(h;t,t) &=& - \ln\frac{h}{\mu^2} - I_2\left(\frac{h}{t}\right) \,.
\end{eqnarray}

In the remainder of this appendix, we list our analytic results for the
threshold corrections $\Delta\overline{r}$ and $\delta_x$ with $x=W,Z,H,t,b$
through two electroweak and three QCD loops.
The pure QCD corrections, of orders $\mathcal{O}(\alpha_s^n)$ with $n=1,2,3$,
only arise for $\delta_t$ and $\delta_b$.
In the case of $\delta_t$, the bottom quark is treated as massless.
In the case of $\delta_b$, the top quark is decoupled.
The $\mathcal{O}(\alpha)$ and $\mathcal{O}(\alpha\alpha_s)$ corrections are
exact, except that the light-fermion masses are neglected.
In the $\mathcal{O}(\alpha\alpha_s)$ corrections, also $M_b=0$ is put, except
in $A_0(b)$.
Our exact formulae for the $\mathcal{O}(\alpha^2)$ corrections are too lengthy
to be presented here, so that we resort to the gaugeless-limit approximation.
The corresponding results 
for $\delta_w$ and $\delta_z$ vanish,
which provides a welcome check for our calculation.
We are thus left with those for $\Delta\overline{r}$, $\delta_h$, $\delta_t$, and
$\delta_b$.
Our master formula reads
\begin{equation}
1+\delta =1+ \delta_\mathrm{QCD} + \frac{g^2}{16\pi^2}  x^{1,0}
          + \frac{g^2}{16\pi^2}\, \frac{g_s^2}{16\pi^2} C_F x^{1,1}
          + \left(\frac{g^2}{16\pi^2}\right)^2 x^{2,0} \,,
\label{eq:master}
\end{equation}
where $\delta=\Delta\overline{r},\delta_x$ and the coefficients $x^{1,0}$, $x^{1,1}$, and
$x^{2,0}$ correspondingly carry the subscripts $\Delta\overline{r}$ and
$x=W,Z,H,t,b$.

\boldmath
\subsection{$\delta_\mathrm{QCD}$}
\label{app:qcd}
\unboldmath

The pure QCD correction in Eq.~(\ref{eq:master}) is given by 
\begin{align}
  \label{eq:mMqcd}
\delta_{\rm QCD}(\mu) &=
  %
  %
  \frac{\alpha_s}{4\pi}\left[-\frac{16}{3} - 4l_{\mu M}\right]
  %
  %
  + \left(\frac{\alpha_s}{4\pi}\right)^2\left[
    -\frac{3305}{18}
    -\frac{64\zeta_2}{3}
    +\frac{8 \zeta_3}{3}
    -\frac{32 \zeta_2}{3}\ln 2\right. \nonumber\\
  & +n_h \left(\frac{143}{9}-\frac{32 \zeta_2}{3}\right)
  + n_l\left(\frac{71}{9}+\frac{16 \zeta_2}{3}\right) + 
  %
  %
  l_{\mu M}\left(-\frac{314}{3} + (n_h+n_l)\frac{52}{9}\right)\nonumber\\
  %
  %
  &\left.+l_{\mu M}^2\left(
      -14+(n_h+n_l)\frac{4}{3}
    \right)
  \right]
  %
  %
  + \left(\frac{\alpha_s}{4\pi}\right)^3\left[
    -\frac{1259285}{162}-\frac{99980\zeta_2}{27}+\frac{584 \zeta_3}{9}\right.\nonumber\\
  &+\frac{4864 a_4}{9}+\frac{6820\zeta_4}{9}-\frac{4928 
    \zeta_2}{3}\ln 2-\frac{13640 \zeta_5}{27}+\frac{10648 \zeta_2 \zeta_3}{9}+\frac{896
    \zeta_2}{9}\ln^2 2\nonumber\\
  &+\frac{608 }{27}\ln^4 2
  +n_h\left(\frac{315526}{243}-\frac{215728
      \zeta_2}{81}-\frac{6008 \zeta_3}{27}-\frac{512 a_4}{27}-\frac{6560
      \zeta_4}{27}\right.\nonumber\\
  &\left.+\frac{81920 
      \zeta_2}{27}\ln 2-\frac{64 }{81}\ln^4 2-80 \zeta_5+96 \zeta_2 \zeta_3+\frac{128
      \zeta_2}{27}\ln^2 2\right)
  +n_l\left(
    \frac{172318}{243}\right.\nonumber\\
  &+\frac{15056 \zeta_2}{27}+\frac{5656
    \zeta_3}{27}-\frac{512 a_4}{27} -\frac{4880 \zeta_4}{27} +\frac{1408
    \zeta_2}{27}\ln 2-\frac{64}{81} 
  \ln^4 2\nonumber\\
  &\left.-\frac{256\zeta_2}{27}  \ln^2 2\right)
  + n_h^2\left(
    -\frac{18962}{729}+\frac{512 \zeta_2}{135}+\frac{352 \zeta_3}{27}
  \right)
  + n_l^2\left(
    -\frac{4706}{729}\right.\nonumber\\
  &\left.-\frac{416 \zeta_2}{27}-\frac{224 \zeta_3}{27}
  \right)
  + n_hn_l\left(
    -\frac{23668}{729}+\frac{416 \zeta_2}{27}+\frac{128 \zeta_3}{27}
  \right)
  %
  %
  + l_{\mu M}\left(-\frac{42650}{9}\right.\nonumber\\
  & \left.-192  \zeta_2\ln 2-384 \zeta_2+48 \zeta_3
    +n_h\left(
      +\frac{18052}{27}-\frac{1472
        \zeta_2}{9}+\frac{448 \zeta_3}{9}+\frac{128 \zeta_2}{9}\ln 2
    \right)\right.\nonumber\\
  &+n_l\left(
    \frac{14164}{27} +\frac{1120
      \zeta_2}{9}+\frac{448 \zeta_3}{9}+ \frac{128  \zeta_2}{9}\ln 2
  \right)
  +n_h^2\left(
    -\frac{1576}{81}+\frac{128 \zeta_2}{9}
  \right)\nonumber\\
  &\left.+n_l^2\left(
      -\frac{712}{81}-\frac{64 \zeta_2}{9}
    \right)+n_hn_l\left(
      -\frac{2288}{81}+\frac{64 \zeta_2}{9}
    \right)\right) +
  %
  %
  l_{\mu M}^2\left(-\frac{3034}{3}+\left(n_h+n_l\right)\frac{428}{3}\right.
\nonumber\\
  &\left.\left.-(n_h+n_l)^2\frac{104}{27}\right)
    +l_{\mu M}^3\left(-81+(n_h+n_l)\frac{128}{9}-(n_h+n_l)^2\frac{16}{27}\right)
  \right]\,,
\end{align}
where $n_l$ is the number of massless quarks, $n_h=1$ is the number of heavy
quarks with pole mass $M$, $l_{\mu M}=\ln(\mu^2/M^2)$, and
$a_4=\li_4(1/2) \approx 0.517479$.


\boldmath
\subsection{$\Delta\overline{r}$}
\label{app:dr}
\unboldmath

The coefficients $x^{1,0}$, $x^{1,1}$, and $x^{2,0}$ in Eq.~(\ref{eq:master})
for $\Delta\overline{r}$ read
\begin{align}
x_{\Delta\overline{r}}^{1,0}&=
        A_0(t)   (
          - 2 \frac{t}{h w} N_c
          + \frac{1}{2} \frac{b}{w (t-b)} N_c
          + \frac{1}{2} \frac{1}{w} N_c
          )
       + A_0(b)   (
          - \frac{1}{2} \frac{b}{w (t-b)} N_c
          - 2 \frac{b}{h w} N_c
          )
\nonumber\\
&
       + A_0(h)   (
           \frac{3}{4} \frac{1}{w}
          - \frac{3}{4} \frac{1}{(h-w)}
          )
       + A_0(z)   (
           \frac{3}{2} \frac{z}{h w}
          + \frac{3}{4} \frac{1}{z S_w^2}
          - \frac{3}{4} \frac{1}{w}
          )
\nonumber\\
&
       + A_0(w)   (
          - \frac{3}{4} \frac{1}{z S_w^2}
          + 3 \frac{1}{h}
          - \frac{3}{2} \frac{1}{w}
          + \frac{3}{4} \frac{1}{(h-w)}
          )
       - \frac{1}{4}
          + \frac{z^2}{h w}
          - \frac{1}{4} \frac{b^2}{w (t-b)} N_c
\nonumber\\
&
          + \frac{1}{4} \frac{t b}{w (t-b)} N_c
          + \frac{1}{4} \frac{t}{w} N_c
          - \frac{1}{8} \frac{h}{w}
          - \frac{1}{8} \frac{z}{w}
          + 2 \frac{w}{h} \,,
\label{eq:x10Dr}\\
x_{\Delta\overline{r}}^{1,1}&=  \frac{N_c}{w} \Bigg\{
        A_0(t)^2   (
           \frac{1}{t}
          - 12 \frac{1}{h}
          )
       + \frac{5}{2} A_0(t)
       + A_{0,\varepsilon}(t)
       - I_0(0,0,t)
       + 16 \frac{t^2}{h}
       - \frac{37}{8} t \Bigg\} \,,
\\
x_{\Delta\overline{r},\mathrm{gl}}^{2,0}&= \frac{1}{w^2} \Bigg\{ 
        A_0^2(t) N_c  \Big(
          - \frac{3}{32} \frac{h}{t}
          + \frac{1}{16}
          + \frac{11}{8} \frac{t}{h}
          \Big)
       + A_0(t) A_0(h) N_c  \Big(
          - \frac{5}{16}
          + \frac{5}{8} \frac{t}{h}
          \Big)
\nonumber\\
&
       + A_0(t) B_0(t;h,t) N_c  \Big(
          - \frac{1}{8} h
          + \frac{3}{2} t
          - 4 \frac{t^2}{h}
          \Big)
       + A_0(t) \re B_0(t;0,0) N_c  \Big(
           \frac{1}{8} t
          - \frac{t^2}{h}
          \Big)
\nonumber\\
&
       + A_0(t) B_0(h;t,t) N_c^2  \Big(
           \frac{t^2}{h}
          - 4 \frac{t^3}{h^2}
          \Big)
       + \frac{9}{4} t N_c A_0(t) B_0(h;h,h)
\nonumber\\
&
       + \frac{3}{4} t N_c A_0(t) \re B_0(h;0,0) 
       + A_0(t) N_c  \Big(
           \frac{9}{16} t
          - \frac{19}{8} \frac{t^2}{h}
          \Big)
       + A_0^2(h)   \Big(
           \frac{27}{32}
          - \frac{15}{32} \frac{t}{h} N_c
\nonumber\\
&
          + \frac{21}{16} \frac{t^2}{h^2} N_c
          \Big)
       + A_0(h) B_0(h;t,t) N_c  \Big(
           \frac{3}{8} t
          - \frac{3}{2} \frac{t^2}{h}
          \Big)
       + \frac{27}{32} h A_0(h) B_0(h;h,h)
\nonumber\\
&
       + \frac{9}{32} h A_0(h) \re B_0(h;0,0)
       + A_0(h)   (
          - \frac{33}{32} h
          + \frac{1}{2} t N_c
          - \frac{5}{4} \frac{t^2}{h} N_c
          )
\nonumber\\
&
       + B_0(t;h,t) N_c  \Big(
          - \frac{3}{16} t h
          + \frac{5}{4} t^2
          - 2 \frac{t^3}{h}
          \Big)
       + \re B_0(t;0,0) N_c  \Big(
           \frac{3}{16} t^2
          - \frac{1}{2} \frac{t^3}{h}
          \Big)
\nonumber\\
&
       + B_0(h;t,t) N_c  \Big(
           \frac{5}{16} t h
          - \frac{5}{4} t^2
          \Big)
       + \frac{45}{64} h^2 B_0(h;h,h)
       + \frac{15}{64} h^2 \re B_0(h;0,0)
\nonumber\\
&
       + I_0(h,t,t) N_c  \Big(
          - \frac{3}{16} h
          - \frac{3}{8} t
          + \frac{9}{4} \frac{t^2}{h}
          \Big)
       + I_0(0,h,t) N_c  \Big(
           \frac{3}{16} h
          - \frac{9}{16} t
          + \frac{3}{8} \frac{t^2}{h}
          \Big)
\nonumber\\
&
       + \frac{131}{128} h^2
          + \frac{1}{4} h^2 \zeta_2
          - \frac{243}{32} h^2 S_2
          - t h N_c
          - \frac{9}{16} t h N_c \zeta_2
          + \frac{59}{64} t^2 N_c
          + \frac{35}{32} t^2 N_c \zeta_2
\nonumber\\
&
          + \frac{77}{16} \frac{t^3}{h} N_c
          + \frac{29}{16} \frac{t^3}{h} N_c \zeta_2
        \Bigg\}\,.
\label{eq:x20Dr}
\end{align}
Equations~(\ref{eq:x10Dr})--(\ref{eq:x20Dr}) agree with Eqs.~(37)--(43) in
Ref.~\cite{Kniehl:2014yia}.


\boldmath
\subsection{$\delta_W$}
\unboldmath

The coefficients $x^{1,0}$ and $x^{1,1}$ in Eq.~(\ref{eq:master}) for
$\delta_W$ read
\begin{align}
x_W^{1,0}&=
        A_0(t)   (
          - \frac{1}{6} \frac{t}{w^2} N_c
          - \frac{1}{2} \frac{b}{w (t-b)} N_c
          + \frac{1}{6} \frac{b}{w^2} N_c
          - \frac{1}{6} \frac{1}{w} N_c
          )
       + A_0(b)   (
           \frac{1}{6} \frac{t}{w^2} N_c
\nonumber\\
&
          + \frac{1}{2} \frac{b}{w (t-b)} N_c
          - \frac{1}{6} \frac{b}{w^2} N_c
          + \frac{1}{3} \frac{1}{w} N_c
          )
       + A_0(h)   (
           \frac{1}{12} \frac{h}{w^2}
          - \frac{1}{4} \frac{1}{w}
          + \frac{3}{4} \frac{1}{(h-w)}
          )
\nonumber\\
&
       + A_0(z)   (
           \frac{1}{12} \frac{z}{w^2}
          - \frac{3}{4} \frac{1}{z S_w^2}
          - 2 \frac{1}{z}
          + \frac{17}{12} \frac{1}{w}
          )
       + A_0(w)   (
          - \frac{1}{12} \frac{h}{w^2}
          - \frac{1}{12} \frac{z}{w^2}
          + \frac{3}{4} \frac{1}{z S_w^2}
\nonumber\\
&
          - 4 \frac{1}{z}
          + \frac{9}{2} \frac{1}{w}
          - \frac{3}{4} \frac{1}{(h-w)}
          )
       + B_0(w;t,b)   (
           \frac{1}{3} N_c
          - \frac{1}{6} \frac{b^2}{w^2} N_c
          + \frac{1}{3} \frac{t b}{w^2} N_c
          - \frac{1}{6} \frac{t^2}{w^2} N_c
\nonumber\\
&
          - \frac{1}{6} \frac{t}{w} N_c
          - \frac{1}{6} \frac{b}{w} N_c
          )
       + B_0(w;w,h)   (
           1
          + \frac{1}{12} \frac{h^2}{w^2}
          - \frac{1}{3} \frac{h}{w}
          )
       + B_0(w;w,z)   (
          - \frac{17}{3}
\nonumber\\
&  
          + \frac{1}{12} \frac{z^2}{w^2}
          + \frac{4}{3} \frac{z}{w}
          - 4 \frac{w}{z}
          )
       + \re B_0(w;0,0)   (
           \frac{1}{3} n_G
          - \frac{1}{3} N_c
          + \frac{1}{3} N_c n_G
          )
       - \frac{247}{36}
          - \frac{1}{9} n_G
\nonumber\\
&
          - \frac{1}{9} N_c n_G
          + \frac{1}{4} \frac{b^2}{w (t-b)} N_c
          - \frac{1}{4} \frac{t b}{w (t-b)} N_c
          + \frac{1}{12} \frac{t}{w} N_c
          + \frac{1}{3} \frac{b}{w} N_c
          - \frac{1}{24} \frac{h}{w}
\nonumber\\
&
          - \frac{1}{24} \frac{z}{w}
          + 4 \frac{w}{z}\,,
\\
x_W^{1,1}&=  N_c \Bigg\{
        A_0^2(t)   (
           \frac{1}{3} \frac{1}{w^2}
          + \frac{1}{t w}
          )
       + A_0(t) B_0(w;0,t)   (
           \frac{4}{3} \frac{1}{t}
          + \frac{5}{3} \frac{1}{w}
          )
       + A_0(t)   (
          - \frac{5}{6} \frac{t}{w^2}
\nonumber\\
&
          - \frac{4}{3} \frac{1}{t}
          - \frac{25}{6} \frac{1}{w}
          )
       + A_{0,\varepsilon}(t)   (
          - \frac{2}{3} \frac{t}{w^2}
          + \frac{4}{3} \frac{1}{t}
          - \frac{1}{3} \frac{1}{w}
          )
       + B_0^2(w;0,t)   (
           \frac{2}{3}
          + \frac{5}{6} \frac{t}{w}
          )
\nonumber\\
&
       + B_0(w;0,t)   (
           1
          - \frac{3}{2} \frac{t^2}{w^2}
          - \frac{5}{2} \frac{t}{w}
          )
       + \re B_0(w;0,0)   (
          - 1
          + n_G
          )
       + \frac{1}{w} I_0(0,0,t)
\nonumber\\
&
       + T_0(w;t,0,0)   (
           \frac{4}{3}
          - \frac{2}{3} \frac{t^2}{w^2}
          + \frac{2}{3} \frac{t}{w}
          )
       + M_0(w;0,0,t,t,0)   (
           \frac{1}{3} \frac{t^3}{w^2}
          - t
          + \frac{2}{3} w 
          ) 
\nonumber\\
&
       + M_0(w;0,0,0,0,0)   (
          - \frac{2}{3} w
          + \frac{2}{3} w n_G
          )
       + \frac{2}{3}
          + \frac{31}{12} n_G
          + \frac{23}{12} \frac{t}{w} \Bigg\} \,.
\end{align}


\boldmath
\subsection{$\delta_Z$}
\unboldmath

The coefficients $x^{1,0}$ and $x^{1,1}$ in Eq.~(\ref{eq:master}) for
$\delta_Z$ read
\begin{align}
x_Z^{1,0} &=
        A_0(t)   (
          - \frac{1}{2} \frac{b}{w (t-b)} N_c
          + \frac{32}{27} \frac{w}{z^2} N_c
          - \frac{40}{27} \frac{1}{z} N_c
          + \frac{7}{54} \frac{1}{w} N_c
          )
       + A_0(b)   (
           \frac{1}{2} \frac{b}{w (t-b)} N_c
\nonumber\\
&
          + \frac{8}{27} \frac{w}{z^2} N_c
          - \frac{4}{27} \frac{1}{z} N_c
          + \frac{5}{27} \frac{1}{w} N_c
          )
       + A_0(h)   (
           \frac{1}{12} \frac{h}{z w}
          - \frac{1}{4} \frac{1}{w}
          + \frac{3}{4} \frac{1}{(h-w)}
          )
\nonumber\\
&
       + A_0(z)   (
          - \frac{1}{12} \frac{h}{z w}
          - \frac{3}{4} \frac{1}{z S_w^2}
          + \frac{11}{12} \frac{1}{w}
          )
       + A_0(w)   (
          - 4 \frac{w}{z^2}
          + \frac{3}{4} \frac{1}{z S_w^2}
          + \frac{4}{3} \frac{1}{z}
          + \frac{5}{3} \frac{1}{w}
\nonumber\\
&
          - \frac{3}{4} \frac{1}{(h-w)}
          )
       + B_0(z;t,t)   (
          - \frac{20}{27} N_c
          + \frac{32}{27} \frac{t w}{z^2} N_c
          - \frac{40}{27} \frac{t}{z} N_c
          + \frac{7}{54} \frac{t}{w} N_c
\nonumber\\
&
          + \frac{17}{54} \frac{z}{w} N_c
          + \frac{16}{27} \frac{w}{z} N_c
          )
       + B_0(z;b,b)   (
          - \frac{2}{27} N_c
          + \frac{8}{27} \frac{b w}{z^2} N_c
          - \frac{4}{27} \frac{b}{z} N_c
          - \frac{17}{54} \frac{b}{w} N_c
\nonumber\\
&
          + \frac{5}{54} \frac{z}{w} N_c
          + \frac{4}{27} \frac{w}{z} N_c
          )
       + B_0(z;z,h)   (
           \frac{1}{12} \frac{h^2}{z w}
          - \frac{1}{3} \frac{h}{w}
          + \frac{z}{w}
          )
       + B_0(z;w,w)   (
           \frac{4}{3}
\nonumber\\
&
          - 4 \frac{w^2}{z^2}
          + \frac{1}{12} \frac{z}{w}
          - \frac{17}{3} \frac{w}{z}
          )
       + \re B_0(z;0,0)   (
          - 2 n_G
          + \frac{22}{27} N_c
          - \frac{22}{27} N_c n_G
          + \frac{z}{w} n_G
\nonumber\\
&
          - \frac{11}{27} \frac{z}{w} N_c
          + \frac{11}{27} \frac{z}{w} N_c n_G
          + \frac{4}{3} \frac{w}{z} n_G
          - \frac{20}{27} \frac{w}{z} N_c
          + \frac{20}{27} \frac{w}{z} N_c n_G
          )
       - \frac{11}{36}
          + \frac{2}{3} n_G
\nonumber\\
&
          + \frac{22}{81} N_c n_G
          - 4 \frac{w^2}{z^2}
          + \frac{8}{27} \frac{b w}{z^2} N_c
          + \frac{1}{4} \frac{b^2}{w (t-b)} N_c
          + \frac{32}{27} \frac{t w}{z^2} N_c
          - \frac{1}{4} \frac{t b}{w (t-b)} N_c
\nonumber\\
&
          - \frac{40}{27} \frac{t}{z} N_c
          + \frac{41}{108} \frac{t}{w} N_c
          - \frac{4}{27} \frac{b}{z} N_c
          + \frac{5}{27} \frac{b}{w} N_c
          - \frac{1}{24} \frac{h}{w}
          + \frac{5}{72} \frac{z}{w}
          - \frac{1}{3} \frac{z}{w} n_G
\nonumber\\
&
          - \frac{11}{81} \frac{z}{w} N_c n_G
          + \frac{4}{3} \frac{w}{z}
          - \frac{4}{9} \frac{w}{z} n_G
          - \frac{20}{81} \frac{w}{z} N_c n_G \,,
\\
x_Z^{1,1} &= N_c \Bigg\{
       - \frac{1}{t w} A_0^2(t)
       + A_0(t) B_0(z;t,t)   (
           \frac{2}{3} \frac{t}{z w}
          - \frac{25}{9} \frac{z}{w (4t-z)}
          - \frac{64}{9} \frac{w}{z (4t-z)}
\nonumber\\
&
          + \frac{128}{27} \frac{w}{z^2}
          - \frac{160}{27} \frac{1}{z}
          + \frac{14}{27} \frac{1}{w}
          + \frac{80}{9} \frac{1}{(4t-z)}
          )
       + A_0(t)   (
           \frac{4}{3} \frac{t}{z w}
          + \frac{25}{9} \frac{z}{w (4t-z)}
\nonumber\\
&
          + \frac{64}{9} \frac{w}{z (4t-z)}
          + \frac{448}{27} \frac{w}{z^2}
          - \frac{560}{27} \frac{1}{z}
          + \frac{287}{54} \frac{1}{w}
          - \frac{80}{9} \frac{1}{(4t-z)}
          )
       + A_{0,\varepsilon}(t)   (
           \frac{2}{3} \frac{t}{z w}
\nonumber\\
&
          - \frac{25}{9} \frac{z}{w (4t-z)}
          - \frac{64}{9} \frac{w}{z (4t-z)}
          + \frac{128}{27} \frac{w}{z^2}
          - \frac{160}{27} \frac{1}{z}
          + \frac{68}{27} \frac{1}{w}
          + \frac{80}{9} \frac{1}{(4t-z)}
          )
\nonumber\\
&
       + B_0(z;t,t)^2   (
           \frac{20}{9}
          - \frac{25}{36} \frac{z^2}{w (4t-z)}
          + \frac{32}{27} \frac{t w}{z^2}
          + \frac{1}{3} \frac{t^2}{z w}
          - \frac{40}{27} \frac{t}{z}
          + \frac{7}{54} \frac{t}{w}
          - \frac{25}{36} \frac{z}{w}
\nonumber\\
&
          + \frac{20}{9} \frac{z}{(4t-z)}
          - \frac{16}{9} \frac{w}{z}
          - \frac{16}{9} \frac{w}{(4t-z)}
          )
       + B_0(z;t,t)   (
          - \frac{100}{9}
          + \frac{25}{12} \frac{z^2}{w (4t-z)}
\nonumber\\
&
          + \frac{128}{9} \frac{t w}{z^2}
          + \frac{2}{3} \frac{t^2}{z w}
          - \frac{160}{9} \frac{t}{z}
          + \frac{29}{9} \frac{t}{w}
          + \frac{143}{36} \frac{z}{w}
          - \frac{20}{3} \frac{z}{(4t-z)}
          + \frac{80}{9} \frac{w}{z}
\nonumber\\
&
          + \frac{16}{3} \frac{w}{(4t-z)}
          )
       + B_{0,\varepsilon}(z;t,t)   (
           \frac{20}{9}
          - \frac{25}{36} \frac{z^2}{w (4t-z)}
          - \frac{64}{27} \frac{t w}{z^2}
          + \frac{80}{27} \frac{t}{z}
          - \frac{7}{27} \frac{t}{w}
\nonumber\\
&
          - \frac{25}{36} \frac{z}{w}
          + \frac{20}{9} \frac{z}{(4t-z)}
          - \frac{16}{9} \frac{w}{z}
          - \frac{16}{9} \frac{w}{(4t-z)}
          )
       + \re B_0(z;0,0)   (
           \frac{20}{9}
          - \frac{22}{9} n_G
\nonumber\\
&
          - \frac{17}{18} \frac{z}{w}
          + \frac{11}{9} \frac{z}{w} n_G
          - \frac{16}{9} \frac{w}{z}
          + \frac{20}{9} \frac{w}{z} n_G
          )
       + \frac{1}{w} I_0(0,0,t)
       + S_0(z;0,t,t)   (
          - \frac{1}{3} \frac{t}{z w}
\nonumber\\
&
          - \frac{32}{9} \frac{w}{z^2}
          + \frac{40}{9} \frac{1}{z}
          - \frac{17}{9} \frac{1}{w}
          )
       + V_0(z;t,t,0,t)   (
          - \frac{128}{27} \frac{t w}{z}
          - \frac{68}{27} \frac{t z}{w}
          + \frac{320}{27} \frac{t^2}{z}
          - \frac{28}{27} \frac{t^2}{w}
\nonumber\\
&
          - \frac{256}{27} \frac{t^2 w}{z^2}
          + \frac{160}{27} \frac{t}{1}
          )
       + M_0(z;t,t,t,t,0)   (
           \frac{17}{27} \frac{z^2}{w}
          - \frac{t z}{w}
          + \frac{160}{27} \frac{t^2}{z}
          - \frac{14}{27} \frac{t^2}{w}
\nonumber\\
&
          - \frac{128}{27} \frac{t^2 w}{z^2}
          - \frac{40}{27} \frac{z}{1}
          + \frac{32}{27} \frac{w}{1}
          )
       + M_0(z;0,0,0,0,0)   (
          - \frac{17}{27} \frac{z^2}{w}
          + \frac{22}{27} \frac{z^2}{w} n_G
          + \frac{40}{27} \frac{z}{1}
\nonumber\\
&
          - \frac{44}{27} \frac{z}{1} n_G
          - \frac{32}{27} \frac{w}{1}
          + \frac{40}{27} \frac{w}{1} n_G
          )
       - \frac{25}{9}
          - \frac{341}{54} n_G
          + \frac{32}{27} \frac{t w}{z^2}
          - \frac{1}{3} \frac{t^2}{z w}
          - \frac{40}{27} \frac{t}{z}
\nonumber\\
&
          - \frac{197}{216} \frac{t}{w}
          + \frac{85}{72} \frac{z}{w}
          + \frac{341}{108} \frac{z}{w} n_G
          + \frac{20}{9} \frac{w}{z}
          + \frac{155}{27} \frac{w}{z} n_G \Bigg\}\,.
\end{align}


\boldmath
\subsection{$\delta_H$}
\unboldmath

The coefficients $x^{1,0}$, $x^{1,1}$, and $x^{2,0}$ in Eq.~(\ref{eq:master})
for $\delta_H$ read
\begin{align}
x_H^{1,0} &=
        A_0(t)   \Big(
          - \frac{1}{2} \frac{b}{w (t-b)} N_c
          - \frac{1}{2} \frac{1}{w} N_c
          \Big)
       + A_0(b) \frac{1}{2} \frac{b}{w (t-b)} N_c
       + A_0(h) \frac{3}{4} \frac{1}{(h-w)}
\nonumber\\
&
       + A_0(z)   \Big(
          - \frac{3}{4} \frac{1}{w S_w^2}
          + \frac{7}{4} \frac{1}{w}
          \Big)
       + A_0(w)   \Big(
           \frac{3}{4} \frac{1}{w S_w^2}
          + \frac{5}{4} \frac{1}{w}
          - \frac{3}{4} \frac{1}{(h-w)}
          \Big)
\nonumber\\
&
       + B_0(h;t,t)   \Big(
          - 2 \frac{t^2}{h w} N_c
          + \frac{1}{2} \frac{t}{w} N_c
          \Big)
       + B_0(h;b,b)   \Big(
          - 2 \frac{b^2}{h w} N_c
          + \frac{1}{2} \frac{b}{w} N_c
          \Big)
\nonumber\\
&
       + \frac{9}{8} \frac{h}{w} B_0(h;h,h)
       + B_0(h;z,z)   \Big(
           \frac{3}{2} \frac{z^2}{h w}
          + \frac{1}{8} \frac{h}{w}
          - \frac{1}{2} \frac{z}{w}
          \Big)
\nonumber\\
&
       + B_0(h;w,w)   \Big(
          - 1
          + \frac{1}{4} \frac{h}{w}
          + 3 \frac{w}{h}
          \Big)
       + \frac{1}{4}
          - 2 \frac{w}{h}
          - \frac{z^2}{w h}
          + \frac{1}{8} \frac{z}{w}
          + \frac{1}{8} \frac{h}{w}
          - \frac{1}{4} \frac{t}{w} N_c
\nonumber\\
&
          + \frac{1}{4} \frac{b^2}{w (t-b)} N_c
          - \frac{1}{4} \frac{t b}{w (t-b)} N_c \,,
\\
x_H^{1,1} &= N_c \Bigg\{
       - \frac{1}{t w} A_0(t)^2 
       + A_0(t) B_0(h;t,t)   (
          - 18 \frac{t}{h w}
          + 3 \frac{1}{w}
          )
       - \frac{5}{2} \frac{1}{w} A_0(t)
\nonumber\\
&
       + A_{0,\varepsilon}(t)   (
           6 \frac{t}{h w}
          - \frac{1}{w}
          )
       + B_0(h;t,t)^2   (
          - 7 \frac{t^2}{h w}
          + \frac{3}{2} \frac{t}{w}
          )
       + B_0(h;t,t)   (
          - 10 \frac{t^2}{h w}
\nonumber\\
&
          + 2 \frac{t}{w}
          )
       + 4 \frac{t^2}{h w} B_{0,\varepsilon}(h;t,t) 
       + \frac{1}{w} I_0(0,0,t)
       - \frac{t}{h w} S_0(h;0,t,t)
\nonumber\\
&
       + V_0(h;t,t,0,t)   (
          - 4 \frac{t^2}{w}
          + 16 \frac{t^3}{h w}
          )
       + M_0(h;t,t,t,t,0)   (
           \frac{t h}{w}
          - 6 \frac{t^2}{w}
          + 8 \frac{t^3}{h w}
          )
\nonumber\\
&
       - 25 \frac{t^2}{h w}
          + \frac{85}{8} \frac{t}{w} \Bigg\} \,,
\label{eq:x11H}\\
x_{H,\mathrm{gl}}^{2,0} &= \frac{1}{w^2} \Bigg\{
        A_0^2(t)   \Big(
           \frac{3}{32} \frac{h}{t} N_c
          - \frac{3}{16} N_c
          + \frac{1}{4} N_c^2
          - \frac{11}{4} \frac{t}{h} N_c
          - 2 \frac{t^2}{h^2} N_c
          + \frac{h N_c}{8 (4t - h)}
          \Big)
\nonumber\\
&
       + A_0(t) A_0(h)   \Big(
           \frac{5}{16} N_c
          - 2 \frac{t}{h} N_c
          - 4 \frac{t^2}{h^2} N_c
          \Big)
       + A_0(t) B_0(t;h,t)   \Big(
           \frac{1}{8} h N_c
          - \frac{1}{4} t N_c
\nonumber\\
&
          - \frac{t^2}{h} N_c
          \Big)
       + A_0(t) \re B_0(t;0,0)   \Big(
          - \frac{1}{8} t N_c
          - \frac{1}{4} \frac{t^2}{h} N_c
          \Big)
       + A_0(t) B_0(h;t,t)   \Big(
          - \frac{1}{16} h N_c
\nonumber\\
&
          + \frac{9}{8} t N_c
          - \frac{1}{2} t N_c^2
          - \frac{11}{4} \frac{t^2}{h} N_c
          + 2 \frac{t^2}{h} N_c^2
          + \frac{h^2 N_c}{16 (4t - h)}
          \Big)
       + A_0(t) B_0(h;h,h)   \Big(
          - \frac{9}{8} h N_c
\nonumber\\
&
          - \frac{9}{4} t N_c
          \Big)
       - \frac{1}{2} t N_c A_0(t) \re B_0(h;0,0)
       + A_0(t)   \Big(
          - \frac{7}{16} h N_c
          + \frac{45}{16} t N_c
          + \frac{1}{4} t N_c^2
\nonumber\\
&
          + \frac{119}{16} \frac{t^2}{h} N_c
          + \frac{9}{2} \frac{t^3}{h^2} N_c
          - \frac{h^2 N_c}{16 (4t - h)}
          \Big)
       + A_{0,\varepsilon}(t)   \Big(
           \frac{1}{4} h N_c
          + \frac{5}{2} t N_c
          - \frac{9}{4} \frac{t^2}{h} N_c
\nonumber\\
&
          + 4 \frac{t^3}{h^2} N_c
          \Big)
       + A_0^2(h)   \Big(
           \frac{99}{64}
          + \frac{15}{32} \frac{t}{h} N_c
          - \frac{21}{16} \frac{t^2}{h^2} N_c
          \Big)
       + A_0(h) B_0(h;t,t)   \Big(
          - \frac{1}{8} t N_c
\nonumber\\
&
          + \frac{t^2}{h} N_c
          - 2 \frac{t^3}{h^2} N_c
          \Big)
       + \frac{81}{32} h A_0(h) B_0(h;h,h)
       + \frac{21}{32} h A_0(h) \re B_0(h;0,0)
\nonumber\\
&
       + A_0(h)   \Big(
          - \frac{9}{16} h
          - \frac{1}{8} t N_c
          - \frac{5}{8} \frac{t^2}{h} N_c
          - 4 \frac{t^3}{h^2} N_c
          \Big)
       + A_{0,\varepsilon}(h)   \Big(
           \frac{3}{32} h
          + \frac{3}{4} t N_c
\nonumber\\
&
          - \frac{5}{8} \frac{t^2}{h} N_c
          + 2 \frac{t^3}{h^2} N_c
          \Big)
       + B_0(t;h,t) B_0(h;t,t)   \Big(
          - \frac{1}{8} t h N_c
          + \frac{7}{4} t^2 N_c
          - 5 \frac{t^3}{h} N_c
          \Big)
\nonumber\\
&
       + B_0(t;h,t)   \Big(
           \frac{3}{16} t h N_c
          - t^2 N_c
          + \frac{t^3}{h} N_c
          \Big)
       + \re B_0(t;0,0) B_0(h;t,t)   \Big(
           \frac{1}{8} t^2 N_c
\nonumber\\
&
          - \frac{5}{4} \frac{t^3}{h} N_c
          \Big)
       + \re B_0(t;0,0)   \Big(
          - \frac{3}{16} t^2 N_c
          + \frac{1}{4} \frac{t^3}{h} N_c
          \Big)
       + B_0^2(h;t,t)   \Big(
           \frac{1}{128} h^2 N_c
\nonumber\\
&
          - \frac{1}{32} t h N_c
          - \frac{1}{16} t^2 N_c
          - \frac{7}{8} \frac{t^3}{h} N_c
          + \frac{h^3 N_c}{128 (4t - h)}
          \Big)
       + B_0(h;t,t) B_0(h;h,h)   \Big(
           \frac{3}{2} t h N_c
\nonumber\\
&
          - \frac{9}{2} t^2 N_c
          \Big)
       + B_0(h;t,t)   \Big(
          - \frac{1}{64} h^2 N_c
          - \frac{1}{4} t h N_c
          + \frac{3}{4} t^2 N_c
          - \frac{1}{4} t^2 N_c^2
\nonumber\\
&
          - \frac{7}{4} \frac{t^3}{h} N_c
          + \frac{t^3}{h} N_c^2
          - \frac{h^3 N_c}{64 (4t - h)}
          \Big)
       + B_{0,\varepsilon}(h;t,t)   \Big(
          + \frac{3}{8} t h N_c
          + \frac{5}{8} t^2 N_c
          - \frac{13}{4} \frac{t^3}{h} N_c
          \Big)
\nonumber\\
&
       + \frac{189}{64} h^2 B_0^2(h;h,h)
       + B_0(h;h,h)   \Big(
           \frac{81}{64} h^2
          + \frac{9}{16} t h N_c
          - \frac{9}{4} t^2 N_c
          \Big)
\nonumber\\
&
       + B_{0,\varepsilon}(h;h,h)   \Big(
           \frac{27}{16} h^2
          + \frac{3}{2} t h N_c
          - \frac{9}{4} t^2 N_c
          \Big)
       + B_0(h;0,t)   (
          - \frac{3}{8} t h N_c
          + \frac{9}{16} t^2 N_c
\nonumber\\
&
          + \frac{5}{16} \frac{t^3}{h} N_c
          - \frac{1}{2} \frac{t^4}{h^2} N_c
          )
       + \re B_0(h;0,0) B_0(h;t,t)   \Big(
           \frac{3}{8} t h N_c
          - \frac{1}{2} t^2 N_c
          \Big)
\nonumber\\
&
       + \frac{27}{16} h^2 \re B_0(h;0,0) B_0(h;h,h)
       + \frac{3}{64} h^2 \Big(\re B_0(h;0,0) \Big)^2 
       + \frac{9}{32} h I_0(h,h,h)
\nonumber\\
&
       + \re B_0(h;0,0)   \Big(
          - \frac{15}{64} h^2
          + \frac{1}{16} t h N_c
          \Big)
\nonumber\\
&
       + I_0(h,t,t)   \Big(
           \frac{3}{16} h N_c
          - \frac{3}{8} t N_c
          + \frac{1}{2} \frac{t^2}{h} N_c
          - 2 \frac{t^3}{h^2} N_c
          \Big)
\nonumber\\
&
       + I_0(0,h,t)   \Big(
          - \frac{3}{16} h N_c
          + \frac{9}{16} t N_c
          - \frac{3}{8} \frac{t^2}{h} N_c
          \Big)
       + I_0(0,0,t)   \Big(
           \frac{7}{8} \frac{t^2}{h} N_c
          - \frac{1}{2} \frac{t^3}{h^2} N_c
          \Big)
\nonumber\\
&
       + \frac{3}{32} h I_0(0,0,h)
       - \frac{17}{8} \frac{t^2}{h} N_c S_0(h;h,t,t)
       + S_0(h;0,t,t)   (
          - \frac{1}{2} t N_c
          - \frac{1}{8} \frac{t^2}{h} N_c
          )
\nonumber\\
&
       + T_0(h;t,0,0)   \Big(
           \frac{1}{4} t h N_c
          - \frac{3}{8} \frac{t^3}{h} N_c
          - \frac{1}{2} \frac{t^4}{h^2} N_c
          \Big)
       + \frac{33}{32} h^2 T_0(h;h,0,0)
\nonumber\\
&
       + U_0(h;t,t,h,t)   \Big(
          - \frac{3}{8} t h N_c
          + 2 \frac{t^3}{h} N_c
          \Big)
       + U_0(h;t,t,0,0)   \Big(
          - \frac{5}{8} t^2 N_c
          + \frac{5}{4} \frac{t^3}{h} N_c
          \Big)
\nonumber\\
&
       + U_0(h;h,h,t,t)   \Big(
          - \frac{3}{2} t h N_c
          + \frac{9}{4} t^2 N_c
          \Big)
       - \frac{27}{32} h^2 U_0(h;h,h,h,h)
\nonumber\\
&
       - \frac{27}{32} h^2 U_0(h;h,h,0,0)
       + M_0(h;t,t,t,t,h)   \Big(
           \frac{1}{8} t^2 h N_c
          + t^3 N_c
          - 2 \frac{t^4}{h} N_c
          \Big)
\nonumber\\
&
       + M_0(h;t,h,t,h,t)   \Big(
           \frac{9}{8} t^2 h N_c
          - 3 t^3 N_c
          \Big)
       - \frac{1}{8} t^2 h N_c M_0(h;t,0,t,0,t)
\nonumber\\
&
       - \frac{1}{4} t^3 N_c M_0(h;t,0,t,0,0)
       + M_0(h;h,t,h,t,t)   \Big(
          + \frac{9}{8} t^2 h N_c
          - 3 t^3 N_c
          \Big)
\nonumber\\
&
       + \frac{81}{32} h^3 M_0(h;h,h,h,h,h)
       + \frac{9}{32} h^3  M_0(h;h,0,h,0,0) 
\nonumber\\
&
       - \frac{1}{8} t^2 h N_c M_0(h;0,t,0,t,t)
       - \frac{1}{4} t^3 N_c M_0(h;0,t,0,t,0)
\nonumber\\
&
       + \frac{9}{32} h^3 M_0(h;0,h,0,h,0)
       + \frac{3}{32} h^3 M_0(h;0,0,0,0,h)
       - \frac{141}{64} h^2
          + \frac{37}{32} h^2 \zeta_2
\nonumber\\
&
          + \frac{243}{32} h^2 S_2
          + \frac{1}{128} h^2 N_c
          + \frac{51}{32} t h N_c
          + \frac{9}{16} t h N_c \zeta_2
          - \frac{11}{16} t^2 N_c
          - \frac{35}{32} t^2 N_c \zeta_2
\nonumber\\
&
          + \frac{1}{16} t^2 N_c^2
          + \frac{19}{16} \frac{t^3}{h} N_c
          - \frac{29}{16} \frac{t^3}{h} N_c \zeta_2
          - \frac{9}{2} \frac{t^4}{h^2} N_c
          + \frac{h^3 N_c}{128 (4t - h)} 
     \Bigg\} \,.
\end{align}
Equation~(\ref{eq:x11H}) agrees with Eq.~(A.35) in Ref.~\cite{Bezrukov:2012sa}.


\boldmath
\subsection{$\delta_t$}
\unboldmath

The coefficients $x^{1,0}$, $x^{1,1}$, and $x^{2,0}$ in Eq.~(\ref{eq:master})
for $\delta_t$ read
\begin{align}
x_t^{1,0} &=
        A_0(t)   (
          - \frac{1}{4} \frac{b}{w (t-b)} N_c
          + \frac{17}{72} \frac{z}{t w}
          - \frac{8}{9} \frac{w}{t z}
          + \frac{7}{9} \frac{1}{t}
          + \frac{1}{4} \frac{1}{w}
          - \frac{1}{4} \frac{1}{w} N_c
          )
\nonumber\\
&
       + A_0(b)   (
           \frac{1}{4} \frac{b}{w (t-b)} N_c
          + \frac{1}{8} \frac{b}{t w}
          + \frac{1}{4} \frac{1}{t}
          + \frac{1}{8} \frac{1}{w}
          )
       + A_0(h)   (
          - \frac{1}{8} \frac{1}{w}
          + \frac{3}{8} \frac{1}{(h-w)}
          )
\nonumber\\
&
       + A_0(z)   (
          - \frac{17}{72} \frac{z}{t w}
          - \frac{4}{9} \frac{w}{t z}
          - \frac{3}{8} \frac{1}{z S_w^2}
          + \frac{5}{9} \frac{1}{t}
          + \frac{3}{8} \frac{1}{w}
          )
       + A_0(w)   (
          - \frac{1}{8} \frac{b}{t w}
          + \frac{3}{8} \frac{1}{z S_w^2}
\nonumber\\
&
          - \frac{1}{4} \frac{1}{t}
          + \frac{7}{8} \frac{1}{w}
          - \frac{3}{8} \frac{1}{(h-w)}
          )
       + B_0(t;h,t)   (
           \frac{1}{2} \frac{t}{w}
          - \frac{1}{8} \frac{h}{w}
          )
       + B_0(t;z,t)   (
           \frac{10}{9}
          - \frac{17}{72} \frac{z^2}{t w}
\nonumber\\
&
          + \frac{5}{9} \frac{z}{t}
          - \frac{7}{72} \frac{z}{w}
          - \frac{4}{9} \frac{w}{t}
          - \frac{8}{9} \frac{w}{z}
          )
       + B_0(t;w,b)   (
           \frac{1}{8}
          + \frac{1}{8} \frac{b^2}{t w}
          + \frac{1}{8} \frac{t}{w}
          + \frac{1}{8} \frac{b}{t}
          - \frac{1}{4} \frac{b}{w}
          - \frac{1}{4} \frac{w}{t}
          )
\nonumber\\
&
       - \frac{9}{8}
          + \frac{1}{8} \frac{b^2}{w (t-b)} N_c
          - \frac{1}{8} \frac{t b}{w (t-b)} N_c
          - \frac{1}{8} \frac{t}{w} N_c
          + \frac{1}{16} \frac{h}{w}
          + \frac{7}{144} \frac{z}{w}
          + \frac{8}{9} \frac{w}{z} \,,
\\
x_t^{1,1} &=
        A_0^2(t)   (
           \frac{17}{12} \frac{z^2}{t^3 w}
          + \frac{3}{4} \frac{h}{t^2 w}
          + \frac{41}{12} \frac{z}{t^2 w}
          - \frac{10}{3} \frac{z}{t^3}
          - 64 \frac{w}{z^2 (4t-z)}
          + 16 \frac{w}{t z^2}
          + \frac{8}{3} \frac{w}{t^3}
\nonumber\\
&
          - 25 \frac{1}{w (4t-z)}
          + 80 \frac{1}{z (4t-z)}
          + \frac{19}{4} \frac{1}{t w}
          - \frac{5}{4} \frac{1}{t w} N_c
          - 20 \frac{1}{t z}
          - \frac{8}{3} \frac{1}{t^2}
          )
\nonumber\\
&
       + A_0(t) A_0(h)   (
          - \frac{13}{4} \frac{1}{h w}
          + \frac{9}{8} \frac{1}{t (h-w)}
          - \frac{9}{8} \frac{1}{t w}
          )
       + A_0(t) A_0(z)   (
          - \frac{85}{24} \frac{z}{t^2 w}
\nonumber\\
&
          + \frac{64}{3} \frac{w}{z^2 (4t-z)}
          - \frac{104}{9} \frac{w}{t z^2}
          - \frac{20}{3} \frac{w}{t^2 z}
          + \frac{25}{3} \frac{1}{w (4t-z)}
          - \frac{80}{3} \frac{1}{z (4t-z)}
          + \frac{3}{4} \frac{1}{z w}
\nonumber\\
&
          - \frac{199}{72} \frac{1}{t w}
          + \frac{130}{9} \frac{1}{t z}
          - \frac{9}{8} \frac{1}{t z S_w^2}
          + \frac{25}{3} \frac{1}{t^2}
          )
       + A_0(t) A_0(w)   (
          - \frac{9}{8} \frac{1}{t (h-w)}
          + \frac{9}{8} \frac{1}{t w}
\nonumber\\
&
          + \frac{9}{8} \frac{1}{t z S_w^2}
          - \frac{5}{4} \frac{1}{t^2}
          )
       + A_0(t) B_0(t;h,t)   (
          - \frac{7}{8} \frac{h}{t w}
          + \frac{7}{2} \frac{1}{w}
          )
       + A_0(t) B_0(t;z,t)   (
          - \frac{119}{72} \frac{z^2}{t^2 w}
\nonumber\\
&
          - \frac{49}{72} \frac{z}{t w}
          + \frac{35}{9} \frac{z}{t^2}
          - \frac{56}{9} \frac{w}{t z}
          - \frac{28}{9} \frac{w}{t^2}
          + \frac{70}{9} \frac{1}{t}
          )
       + A_0(t) \re B_0(t;0,w)   (
          - \frac{1}{2} \frac{w}{t^2}
          + \frac{1}{t}
          + \frac{1}{w}
          )
\nonumber\\
&
       + A_0(t)   (
           \frac{17}{18} \frac{z^2}{t^2 w}
          - \frac{1}{16} \frac{h}{t w}
          - \frac{25}{2} \frac{z}{w (4t-z)}
          + \frac{265}{144} \frac{z}{t w}
          - \frac{20}{9} \frac{z}{t^2}
          - 32 \frac{w}{z (4t-z)}
\nonumber\\
&
          - \frac{80}{9} \frac{w}{t z}
          - \frac{11}{9} \frac{w}{t^2}
          + \frac{629}{72} \frac{1}{t}
          + \frac{69}{8} \frac{1}{w}
          - \frac{11}{8} \frac{1}{w} N_c
          + 40 \frac{1}{(4t-z)}
          )
       + A_{0,\varepsilon}(t)   (
          - \frac{17}{18} \frac{z^2}{t^2 w}
\nonumber\\
&
          - \frac{5}{4} \frac{h}{t w}
          + \frac{25}{6} \frac{z}{w (4t-z)}
          - \frac{23}{12} \frac{z}{t w}
          + \frac{20}{9} \frac{z}{t^2}
          + \frac{32}{3} \frac{w}{z (4t-z)}
          + \frac{8}{3} \frac{w}{t z}
          - \frac{37}{36} \frac{w}{t^2}
          + \frac{17}{8} \frac{1}{t}
\nonumber\\
&
          + \frac{59}{8} \frac{1}{w}
          - \frac{1}{2} \frac{1}{w} N_c
          - \frac{40}{3} \frac{1}{(4t-z)}
          )
       - \frac{1}{h w} A_0^2(h)
       - \frac{1}{2} \frac{1}{w} A_0(h) B_0(t;h,t) 
\nonumber\\
&
       + A_0(h)   (
          - \frac{1}{8} \frac{h}{t w}
          + 3 \frac{1}{w}
          - \frac{3}{8} \frac{1}{(h-w)}
          )
       + A_{0,\varepsilon}(h)   (
           \frac{13}{4} \frac{t}{h w}
          + \frac{9}{4} \frac{1}{w}
          )
\nonumber\\
&
       + A_0^2(z)   (
           \frac{64}{3} \frac{w}{z^2 (4t-z)}
          - \frac{32}{9} \frac{w}{t z^2}
          + \frac{25}{3} \frac{1}{w (4t-z)}
          - \frac{80}{3} \frac{1}{z (4t-z)}
          - \frac{17}{9} \frac{1}{t w}
          + \frac{40}{9} \frac{1}{t z}
          )
\nonumber\\
&
       + A_0(z) B_0(t;z,t)   (
           \frac{25}{6} \frac{z}{w (4t-z)}
          - \frac{17}{18} \frac{z}{t w}
          + \frac{32}{3} \frac{w}{z (4t-z)}
          - \frac{16}{9} \frac{w}{t z}
          + \frac{20}{9} \frac{1}{t}
\nonumber\\
&
          - \frac{40}{3} \frac{1}{(4t-z)}
          )
       + A_0(z)   (
           \frac{17}{36} \frac{z^2}{t^2 w}
          - \frac{125}{12} \frac{z}{w (4t-z)}
          + \frac{77}{24} \frac{z}{t w}
          - \frac{10}{9} \frac{z}{t^2}
          - \frac{80}{3} \frac{w}{z (4t-z)}
\nonumber\\
&
          + \frac{16}{3} \frac{w}{t z}
          + \frac{8}{9} \frac{w}{t^2}
          + \frac{3}{8} \frac{1}{z S_w^2}
          - \frac{20}{3} \frac{1}{t}
          - \frac{3}{4} \frac{1}{w}
          + \frac{100}{3} \frac{1}{(4t-z)}
          )
       + A_{0,\varepsilon}(z)   (
          - \frac{3}{4} \frac{t}{z w}
\nonumber\\
&
          - \frac{175}{12} \frac{z}{w (4t-z)}
          + \frac{17}{3} \frac{z}{t w}
          - \frac{112}{3} \frac{w}{z (4t-z)}
          + \frac{56}{9} \frac{w}{z^2}
          + \frac{32}{3} \frac{w}{t z}
          - \frac{40}{3} \frac{1}{t}
          - \frac{70}{9} \frac{1}{z}
          + \frac{65}{36} \frac{1}{w}
\nonumber\\
&
          + \frac{140}{3} \frac{1}{(4t-z)}
          )
       + A_0^2(w)   (
           \frac{5}{8} \frac{1}{w^2}
          + \frac{5}{4} \frac{1}{t w}
          )
       + A_0(w) \re B_0(t;0,w)   (
           \frac{5}{8} \frac{t}{w^2}
          + \frac{5}{4} \frac{1}{t}
          + \frac{15}{8} \frac{1}{w}
          )
\nonumber\\
&
       + A_0(w)   (
          - \frac{5}{8} \frac{t}{w^2}
          - \frac{1}{4} \frac{w}{t^2}
          - \frac{3}{8} \frac{1}{z S_w^2}
          + \frac{11}{8} \frac{1}{t}
          - 2 \frac{1}{w}
          + \frac{3}{8} \frac{1}{(h-w)}
          )
       + A_{0,\varepsilon}(w)   (
           \frac{5}{8} \frac{t}{w^2}
          - \frac{3}{4} \frac{1}{t}
\nonumber\\
&
          + \frac{17}{8} \frac{1}{w}
          )
       + B_0(t;h,t)   (
           \frac{1}{8} \frac{h^2}{t w}
          + 13 \frac{t}{w}
          - \frac{17}{4} \frac{h}{w}
          )
       + B_{0,\varepsilon}(t;h,t)   (
          - 2 \frac{t}{w}
          + \frac{h}{w}
          )
\nonumber\\
&
       + B_0(t;z,t)   (
           \frac{110}{9}
          - \frac{8}{9} \frac{z w}{t^2}
          + \frac{25}{6} \frac{z^2}{w (4t-z)}
          - \frac{95}{72} \frac{z^2}{t w}
          + \frac{10}{9} \frac{z^2}{t^2}
          - \frac{17}{36} \frac{z^3}{t^2 w}
          - \frac{3}{2} \frac{t}{w}
\nonumber\\
&
          + \frac{20}{9} \frac{z}{t}
          + \frac{1}{18} \frac{z}{w}
          - \frac{40}{3} \frac{z}{(4t-z)}
          - \frac{16}{9} \frac{w}{t}
          - \frac{88}{9} \frac{w}{z}
          + \frac{32}{3} \frac{w}{(4t-z)}
          )
\nonumber\\
&
       + B_{0,\varepsilon}(t;z,t)   (
          - \frac{40}{9}
          - \frac{25}{6} \frac{z^2}{w (4t-z)}
          + \frac{17}{9} \frac{z^2}{t w}
          - \frac{40}{9} \frac{z}{t}
          + \frac{7}{18} \frac{z}{w}
          + \frac{40}{3} \frac{z}{(4t-z)}
\nonumber\\
&
          + \frac{32}{9} \frac{w}{t}
          + \frac{32}{9} \frac{w}{z}
          - \frac{32}{3} \frac{w}{(4t-z)}
          )
       + \re  [B_0^2(t;0,w)]   (
           \frac{5}{4}
          + \frac{5}{8} \frac{t}{w}
          )
       + \re B_0(t;0,w)   (
          - \frac{1}{4}
\nonumber\\
&
          + \frac{1}{4} \frac{w^2}{t^2}
          + \frac{5}{8} \frac{t}{w}
          - \frac{47}{8} \frac{w}{t}
          )
       - \frac{3}{2} \frac{w}{t} \re B_{0,\varepsilon}(t;0,w)
       + I_0(h,t,t)   (
           \frac{1}{8} \frac{h}{t w}
          - \frac{11}{4} \frac{1}{w}
          )
\nonumber\\
&
       + I_0(z,t,t)   (
          - \frac{17}{36} \frac{z^2}{t^2 w}
          + \frac{25}{4} \frac{z}{w (4t-z)}
          - \frac{95}{72} \frac{z}{t w}
          + \frac{10}{9} \frac{z}{t^2}
          + 16 \frac{w}{z (4t-z)}
          - \frac{16}{9} \frac{w}{t z}
\nonumber\\
&
          - \frac{8}{9} \frac{w}{t^2}
          + \frac{20}{9} \frac{1}{t}
          - 20 \frac{1}{(4t-z)}
          )
       + I_0(0,w,t)   (
           \frac{1}{4} \frac{w}{t^2}
          - \frac{13}{8} \frac{1}{t}
          - \frac{13}{8} \frac{1}{w}
          )
\nonumber\\
&
       + \frac{1}{2} \frac{1}{w} N_c I_0(0,0,t) 
       + T_0(t;t,0,0)   (
          - \frac{71}{18}
          - \frac{17}{9} \frac{z^2}{t w}
          + \frac{9}{4} \frac{t}{w}
          - \frac{h}{w}
          + \frac{40}{9} \frac{z}{t}
          - \frac{7}{9} \frac{z}{w}
\nonumber\\
&
          - \frac{23}{9} \frac{w}{t}
          + \frac{56}{9} \frac{w}{z}
          )
       + T_0(t;h,t,0)   (
           \frac{1}{8} \frac{h^2}{t w}
          + \frac{13}{4} \frac{t}{w}
          - \frac{3}{2} \frac{h}{w}
          )
       + T_0(t;z,t,0)   (
          - \frac{70}{9}
\nonumber\\
&
          - \frac{8}{9} \frac{z w}{t^2}
          + \frac{59}{24} \frac{z^2}{t w}
          + \frac{10}{9} \frac{z^2}{t^2}
          - \frac{17}{36} \frac{z^3}{t^2 w}
          - \frac{3}{4} \frac{t}{w}
          - \frac{20}{3} \frac{z}{t}
          + \frac{65}{36} \frac{z}{w}
          + \frac{16}{3} \frac{w}{t}
          + \frac{56}{9} \frac{w}{z}
          )
\nonumber\\
&
       + T_0(t;w,0,0)   (
           \frac{1}{2}
          + \frac{1}{4} \frac{w^2}{t^2}
          + \frac{5}{8} \frac{t}{w}
          - \frac{19}{8} \frac{w}{t}
          )
       + \frac{3}{2} \frac{w}{t} U_0(t;0,w,0,t)
\nonumber\\
&
       + V_0(t;t,h,0,h)   (
           \frac{h^2}{w}
          - 4 \frac{t h}{w}
          )
       + V_0(t;t,z,0,z)   (
           \frac{32}{9} \frac{z w}{t}
          - \frac{40}{9} \frac{z^2}{t}
          + \frac{7}{9} \frac{z^2}{w}
          + \frac{17}{9} \frac{z^3}{t w}
\nonumber\\
&
          - \frac{80}{9} \frac{z}{1}
          + \frac{64}{9} \frac{w}{1}
          )
       + M_0(t;0,t,t,h,t)   (
          - \frac{1}{2} \frac{h^2}{w}
          + 3 \frac{t h}{w}
          - 4 \frac{t^2}{w}
          )
\nonumber\\
&
       + M_0(t;0,t,t,z,t)   (
          - \frac{16}{9} \frac{z w}{t}
          + \frac{20}{9} \frac{z^2}{t}
          + \frac{3}{2} \frac{z^2}{w}
          - \frac{17}{18} \frac{z^3}{t w}
          + \frac{64}{9} \frac{t w}{z}
          + \frac{7}{9} \frac{t z}{w}
          - \frac{80}{9} \frac{t}{1}
          )
\nonumber\\
&
       + M_0(t;0,t,t,0,t)   (
          - \frac{64}{9} \frac{t w}{z}
          + \frac{64}{9} \frac{t}{1}
          )
       + M_0(t;0,0,t,w,0)   (
          - \frac{1}{2} \frac{w^2}{t}
          - \frac{1}{4} \frac{t^2}{w}
          + \frac{3}{4} \frac{w}{1}
          )
\nonumber\\
&
       + M_0(t;0,0,w,t,0)   (
          - \frac{1}{2} \frac{w^2}{t}
          - \frac{1}{4} \frac{t^2}{w}
          + \frac{3}{4} \frac{w}{1}
          )
       - \frac{3299}{144}
          + \frac{1}{4} \frac{w^2}{t^2}
          - \frac{8}{9} \frac{z w}{t^2}
\nonumber\\
&
          + \frac{325}{24} \frac{z^2}{w (4t-z)}
          - \frac{163}{72} \frac{z^2}{t w}
          + \frac{10}{9} \frac{z^2}{t^2}
          - \frac{17}{36} \frac{z^3}{t^2 w}
          + \frac{1}{8} \frac{h^2}{t w}
          - \frac{29}{16} \frac{t}{w}
          + \frac{39}{16} \frac{t}{w} N_c
\nonumber\\
&
          - \frac{53}{16} \frac{h}{w}
          + \frac{40}{9} \frac{z}{t}
          - \frac{53}{96} \frac{z}{w}
          - \frac{130}{3} \frac{z}{(4t-z)}
          - \frac{211}{72} \frac{w}{t}
          + \frac{176}{9} \frac{w}{z}
          + \frac{104}{3} \frac{w}{(4t-z)}
\nonumber\\
&
          - \frac{15}{2} \frac{w^2}{t^2} \zeta_2
          + \frac{45}{4} \frac{w}{t} \zeta_2
          - \frac{15}{4} \frac{t}{w} \zeta_2  \,,
\label{eq:x11t}\\
x^{2,0}_{t,\mathrm{gl}} &=\frac{1}{w^2} \Bigg\{
              A^2_0(h)       \left(
             \frac{1}{128}
             + \frac{1}{4} \frac{t}{h}
             \right)
             + A_0(h) A_0(t)   \left(
             - \frac{3}{32}
             - \frac{13}{32} \frac{t}{h}
             + \frac{9}{4} \frac{t}{h} N_c
             - \frac{1}{2} N_c
             - \frac{1}{32} \frac{h}{t}
             \right)
             \nonumber\\
           & + \frac{9}{64} h A_0(h) B_0(h;h,h)   
             + A_0(h) B_0(h;t,t)   \left(
             \frac{1}{16} t N_c - \frac{1}{4} \frac{t^2}{h} N_c
             \right)
             + \frac{3}{64} h A_0(h) B_0(h;0,0) 
             \nonumber\\
           & + A_0(h) B_0(t;h,t)   \left(
             \frac{1}{2} \frac{t^2}{h}
             - \frac{1}{32} h
             \right)
             + A_0(h) B_0(t;0,0)   \left(
             \frac{1}{16} \frac{t^2}{h}
             + \frac{3}{64} t
             \right)
             \nonumber\\
           & + A_0(h)   \left(
             - \frac{9}{16} \frac{t^2}{h}
             - \frac{1}{2} \frac{t^2}{h} N_c
             - \frac{87}{128} t
             - \frac{5}{16} t N_c
             + \frac{9}{32} h
             + \frac{3}{32} h N_c
             + \frac{1}{32} \frac{h^2}{t}
             \right)
             \nonumber\\
           & + A_{0,\varepsilon}(h)   \left(
              \frac{31}{32} \frac{t^2}{h}
             - \frac{9}{4} \frac{t^2}{h} N_c
             - \frac{23}{64} t
             + \frac{9}{16} t N_c
             + \frac{17}{64} h
             \right)
             \nonumber\\
           & + A^2_0(t)   \left(
             + \frac{1}{4}
             - \frac{1}{2} \frac{t^2}{h^2} N_c^2
             - \frac{3}{2} \frac{t}{h} N_c
             + \frac{1}{2} \frac{t}{h} N_c^2
             + \frac{1}{4} N_c
             - \frac{3}{32} \frac{h}{t}
             \right)
             \nonumber\\
           & + A_0(t) B_0(h;h,h)   \left(
             \frac{9}{8} t N_c
             + \frac{9}{32} h
             \right)
             + A_0(t) B_0(h;t,t)   \Bigg(
             - 2 \frac{t^3}{h^2} N_c^2
             - \frac{1}{2} \frac{t^2}{h} N_c
             \nonumber\\
             &
             + \frac{1}{2} \frac{t^2}{h} N_c^2
             + \frac{1}{8} t N_c
             \Bigg)
             + A_0(t) B_0(h;0,0)   \left(
             \frac{3}{8} t N_c
             + \frac{3}{32} h
             \right)
             \nonumber\\
           & 
             + A_0(t) B_0(t;h,t)   \left(
             - 3 \frac{t^2}{h} N_c
             + \frac{3}{8} t
             + \frac{5}{4} t N_c
             - \frac{11}{32} h
             - \frac{1}{8} h N_c
             \right)
             \nonumber\\
           &+ A_0(t) B_0(t;0,0)   \left(
             - \frac{5}{8} \frac{t^2}{h} N_c
             + \frac{3}{16} t
             \right)
             \nonumber\\
           & 
             + A_0(t)   \left(
             - \frac{3}{4} \frac{t^2}{h} N_c
             + \frac{1}{4} \frac{t^2}{h} N_c^2
             + \frac{117}{128} t
             - \frac{23}{16} t N_c
             + \frac{13}{32} h
             + \frac{3}{16} h N_c
             + \frac{1}{32} \frac{h^2}{t}
             \right)
             \nonumber\\
           & 
             + A_{0,\varepsilon}(t)   \left(
             - 2 \frac{t^2}{h} N_c
             + \frac{3}{16} t
             - \frac{1}{16} t N_c
             + \frac{13}{32} h
             + \frac{3}{16} h N_c
             + \frac{1}{32} \frac{h^2}{t}
             \right)
             \nonumber\\
           & + B_0(h;h,h) B_0(t;h,t)   \left(
               \frac{9}{32} h t
             - \frac{9}{32} h^2
             \right)
             + B_0(h;h,h)   \left(
             - \frac{9}{32} h t
             + \frac{27}{64} h^2
             \right)
             \nonumber\\
           & + B_0(h;t,t) B_0(t;h,t)   \left(
             - \frac{1}{2} \frac{t^3}{h} N_c
             + \frac{5}{8} t^2 N_c
             - \frac{1}{8} h t N_c
             \right)
             \nonumber\\
           &
             + B_0(h;t,t)   \left(
               \frac{1}{2} \frac{t^3}{h} N_c
             - \frac{7}{8} t^2 N_c
             + \frac{3}{16} h t N_c
             \right)
             + B_0(h;0,0) B_0(t;h,t)   \left(
               \frac{3}{32} h t
             - \frac{3}{32} h^2
             \right)
             \nonumber\\
           & 
             + B_0(h;0,0)   \left(
             - \frac{3}{32} h t
             + \frac{9}{64} h^2
             \right)
             + B_0^2(t;h,t)   \left(
               \frac{3}{4} t^2
             - \frac{1}{4} h t
             + \frac{1}{32} h^2
             \right)
             \nonumber\\
           & 
             + B_0(t;h,t)   \left(
             - \frac{3}{2} \frac{t^3}{h} N_c
             + \frac{27}{16} t^2
             - \frac{7}{4} t^2 N_c
             + \frac{37}{64} h t
             + \frac{29}{32} h t N_c
             - \frac{3}{16} h^2
             - \frac{3}{32} h^2 N_c
             \right)
             \nonumber\\
           & 
             + B_{0,\varepsilon}(t;h,t)   \left(
             \frac{13}{32} t^2
               \frac{35}{64} h t
             - \frac{21}{64} h^2
             \right)
             \nonumber\\
           & + B_0(t;0,h)   \left(
             - \frac{3}{128} t^2
             - \frac{3}{128} h t
             + \frac{5}{64} h^2
             - \frac{1}{32} \frac{h^3}{t}
             \right)
             + \frac{11}{32} t^2 B_0(t;0,0) B_0(A(t),h,t)  
             \nonumber\\
           & + \frac{3}{128} t^2 B_0(t;0,0)^2 
             + B_0(t;0,0)   \left(
             - \frac{1}{4} \frac{t^3}{h} N_c
             + \frac{3}{128} t^2
             \right)
             - \frac{3}{64} h I_0(h,h,h) 
             \nonumber\\
           & + I_0(h,t,t)   \left(
             + \frac{3}{4} \frac{t^2}{h} N_c
             - \frac{3}{16} t
             + \frac{3}{16} t N_c
             - \frac{1}{8} h
             - \frac{3}{32} h N_c
             \right)
             \nonumber\\
           & 
             + I_0(0,0,h)   \left(
             - \frac{5}{64} h
             - \frac{1}{32} \frac{h^2}{t}
             \right)
             + I_0(0,0,t)   \left(
             + \frac{1}{2} \frac{t^2}{h} N_c
             - \frac{1}{32} t
             + \frac{1}{16} t N_c
             \right)
             \nonumber\\
           & 
             + \frac{17}{64} t S_0(t;h,h,t) 
             + T_0(t;h,t,0)   \left(
             \frac{29}{32} t^2
             - \frac{3}{2} t^2 N_c
             - \frac{1}{16} h t
             + \frac{3}{4} h t N_c
             - \frac{5}{64} h^2
             \right.
             \nonumber\\
           & 
             \left.
             - \frac{3}{32} h^2 N_c
             \right)
             + T_0(t;h,0,0)   \left(
             \frac{1}{16} t^2
             + \frac{3}{64} h t
             - \frac{1}{32} \frac{h^3}{t}
             \right)
             \nonumber\\
           & + T_0(t;t,0,0)   \left(
             \frac{3}{64} t^2
             + \frac{3}{8} t^2 N_c
             + \frac{5}{32} h t
             \right)
             + U_0(t;h,t,h,t)   \left(
             - \frac{1}{4} t^2
             + \frac{3}{64} h^2
             \right)
             \nonumber\\
           & + U_0(t;h,t,0,0)   \left(
             - \frac{5}{32} t^2
             + \frac{5}{64} h t
             \right)
             + U_0(t;t,h,h,h)   \left(
             - \frac{9}{32} h t
             + \frac{3}{16} h^2
             \right)
             \nonumber\\
           & + U_0(t;t,h,0,0)   \left(
             \frac{3}{32} h^2
             - \frac{11}{32} h t
             \right)
             + \frac{1}{32} h^2 U_0(t;t,0,0,h)  
             \nonumber\\
           & 
             + M_0(t;h,h,t,t,h)   \left(
             \frac{3}{4} h t^2
             - \frac{9}{32} h^2 t
             \right)
             \nonumber\\
           & + M_0(t;h,t,t,h,t)   \left(
              \frac{1}{2} t^3
             - \frac{1}{4} h t^2
             - \frac{1}{32} h^2 t
             \right)
             + \frac{1}{32} h^2 t M_0(t;h,0,t,t,0)
             \nonumber\\
           & + \frac{1}{16} h t^2 M_0(t;h,0,t,0,0)
             + \frac{1}{32} h^2 t M_0(t;0,h,t,t,0)
             + \frac{1}{16} h t^2 M_0(t;0,h,0,t,0)
             \nonumber\\
           & + \frac{1}{32} h^2 t M_0(t;0,0,t,t,h) 
             + \frac{5}{4} \frac{t^3}{h} N_c
             - \frac{197}{512} t^2
             - \frac{23}{128} t^2 N_c
             + \frac{17}{64} h t
             + \frac{3}{8} h t N_c
             \nonumber\\
           & - \frac{21}{64} h^2
             - \frac{3}{32} h^2 N_c
             - \frac{1}{32} \frac{h^3}{t}
             \Bigg\} \,.
\label{eq:x20t}
\end{align}
Equation~(\ref{eq:x11t}) agrees with Eq.~(A.34) in Ref.~\cite{Bezrukov:2012sa}.
An expansion of Eq.~(\ref{eq:x20t}) in $\Delta_H=1-M_H^2/M_t^2$ may be found in
Eqs.~(19)--(23) of Ref.~\cite{Kniehl:2014yia}.


\boldmath
\subsection{$\delta_b$}
\unboldmath

The coefficients $x^{1,0}$, $x^{1,1}$, and $x^{2,0}$ in Eq.~(\ref{eq:master})
for $\delta_b$ read
\begin{align}
x_b^{1,0} &=
        A_0(t)   (
           \frac{1}{8} \frac{t}{b w}
          - \frac{1}{4} \frac{b}{w (t-b)} N_c
          + \frac{1}{4} \frac{1}{b}
          + \frac{1}{8} \frac{1}{w}
          - \frac{1}{4} \frac{1}{w} N_c
          )
       + A_0(b)   (
           \frac{1}{4} \frac{b}{w (t-b)} N_c
\nonumber\\
&
          + \frac{5}{72} \frac{z}{b w}
          - \frac{2}{9} \frac{w}{b z}
          + \frac{5}{18} \frac{1}{b}
          + \frac{1}{4} \frac{1}{w}
          )
       + A_0(h)   (
          - \frac{1}{8} \frac{1}{w}
          + \frac{3}{8} \frac{1}{(h-w)}
          )
       + A_0(z)   (
          - \frac{5}{72} \frac{z}{b w}
\nonumber\\
&
          - \frac{1}{9} \frac{w}{b z}
          - \frac{3}{8} \frac{1}{z S_w^2}
          + \frac{1}{18} \frac{1}{b}
          + \frac{3}{8} \frac{1}{w}
          )
       + A_0(w)   (
          - \frac{1}{8} \frac{t}{b w}
          + \frac{3}{8} \frac{1}{z S_w^2}
          - \frac{1}{4} \frac{1}{b}
          + \frac{7}{8} \frac{1}{w}
\nonumber\\
&
          - \frac{3}{8} \frac{1}{(h-w)}
          )
       + B_0(b;h,b)   (
           \frac{1}{2} \frac{b}{w}
          - \frac{1}{8} \frac{h}{w}
          )
       + B_0(b;z,b)   (
           \frac{1}{9}
          - \frac{5}{72} \frac{z^2}{b w}
          + \frac{1}{18} \frac{z}{b}
\nonumber\\
&
          + \frac{17}{72} \frac{z}{w}
          - \frac{1}{9} \frac{w}{b}
          - \frac{2}{9} \frac{w}{z}
          )
       + B_0(b;w,t)   (
           \frac{1}{8}
          + \frac{1}{8} \frac{t^2}{b w}
          + \frac{1}{8} \frac{t}{b}
          - \frac{1}{4} \frac{t}{w}
          + \frac{1}{8} \frac{b}{w}
          - \frac{1}{4} \frac{w}{b}
          )
\nonumber\\
&
       - \frac{7}{24}
          + \frac{1}{8} \frac{b^2}{w (t-b)} N_c
          - \frac{1}{8} \frac{t b}{w (t-b)} N_c
          - \frac{1}{8} \frac{t}{w} N_c
          + \frac{1}{16} \frac{h}{w}
          - \frac{17}{144} \frac{z}{w}
          + \frac{2}{9} \frac{w}{z} \,,
\label{eq:x10b}
\\
x_b^{1,1} &=
        A_0^2(t)   (
          - \frac{9}{8} \frac{1}{(t-w)^2}
          - \frac{3}{8} \frac{1}{w (t-w)}
          - \frac{1}{2} \frac{1}{t w} N_c
          )
       + A_0(t) A_0(b)   (
          - \frac{9}{8} \frac{w}{b (t-w)^2}
\nonumber\\
&
          + \frac{9}{8} \frac{1}{b w}
          - \frac{3}{4} \frac{1}{b w} N_c
          )
       + A_0(t) A_0(w)   (
          - \frac{9}{4} \frac{w}{(t-w)^3}
          - 3 \frac{1}{(t-w)^2}
          - 3 \frac{1}{w (t-w)}
          )
\nonumber\\
&
       + A_0(t)   (
           \frac{9}{4} \frac{w^2}{(t-w)^3}
          + \frac{33}{4} \frac{w}{(t-w)^2}
          + \frac{39}{8} \frac{1}{w}
          - \frac{1}{w} N_c
          + \frac{27}{4} \frac{1}{(t-w)}
          )
\nonumber\\
&
       + A_{0,\varepsilon}(t)   (
           \frac{9}{4} \frac{w^2}{(t-w)^3}
          + \frac{21}{4} \frac{w}{(t-w)^2}
          + 3 \frac{1}{w}
          - \frac{1}{2} \frac{1}{w} N_c
          + 6 \frac{1}{(t-w)}
          )
\nonumber\\
&
       + A_0^2(b)   (
          - \frac{8}{3} \frac{w}{b^2 z}
          + \frac{8}{3} \frac{1}{b^2}
          )
       + \frac{9}{8} \frac{1}{b (h-w)} A_0(b) A_0(h)
       + A_0(b) A_0(z)   (
           \frac{w}{b z^2}
          + \frac{7}{4} \frac{1}{b w}
\nonumber\\
&
          - \frac{1}{2} \frac{1}{b z}
          - \frac{9}{8} \frac{1}{b w S_w^2}
          )
       + A_0(b) A_0(w)   (
           \frac{9}{8} \frac{w}{b (t-w)^2}
          + \frac{9}{8} \frac{1}{b (t-w)}
          - \frac{9}{8} \frac{1}{b (h-w)}
\nonumber\\
&
          + \frac{9}{8} \frac{1}{b w}
          + \frac{9}{8} \frac{1}{b w S_w^2}
          )
       + A_0(b)   (
           \frac{3}{16} \frac{t}{b w}
          - \frac{3}{8} \frac{t}{b w} N_c
          + \frac{3}{16} \frac{h}{b w}
          - \frac{11}{24} \frac{z}{b w}
          + \frac{9}{8} \frac{w}{b (t-w)}
\nonumber\\
&
          - \frac{17}{6} \frac{w}{b z}
          + \frac{79}{24} \frac{1}{b}
          )
       + A_{0,\varepsilon}(b)   (
           \frac{10}{3} \frac{w}{b z}
          - \frac{10}{3} \frac{1}{b}
          )
       - \frac{3}{8} \frac{1}{(h-w)} A_0(h) 
\nonumber\\
&
       + A_0(z)   (
           \frac{3}{8} \frac{1}{w S_w^2}
          - \frac{15}{8} \frac{1}{w}
          )
       + A_0(w)   (
           \frac{9}{4} \frac{w^2}{(t-w)^3}
          + \frac{9}{4} \frac{w}{(t-w)^2}
          - \frac{3}{8} \frac{1}{w S_w^2}
          - \frac{3}{8} \frac{1}{w}
\nonumber\\
&
          + \frac{3}{8} \frac{1}{(h-w)}
          + \frac{9}{4} \frac{1}{(t-w)}
          )
       + A_{0,\varepsilon}(w)   (
           \frac{9}{4} \frac{w^2}{(t-w)^3}
          + \frac{21}{4} \frac{w}{(t-w)^2}
          + 3 \frac{1}{w}
\nonumber\\
&
          + 6 \frac{1}{(t-w)}
          )
       + I_0(0,w,t)   (
          - \frac{9}{4} \frac{w^2}{(t-w)^3}
          - \frac{21}{4} \frac{w}{(t-w)^2}
          - 3 \frac{1}{w}
          - 6 \frac{1}{(t-w)}
          )
\nonumber\\
&
       + \frac{1}{2} \frac{1}{w} N_c I_0(0,0,t)
       + T_0(b;b,0,0)   (
          - \frac{10}{3}
          + \frac{10}{3} \frac{w}{z}
          )
       - \frac{315}{16}
          - \frac{51}{4} \frac{w^2}{(t-w)^2}
\nonumber\\
&
          - \frac{9}{2} \frac{w^3}{(t-w)^3}
          - \frac{103}{16} \frac{t}{w}
          + \frac{39}{16} \frac{t}{w} N_c
          - \frac{1}{16} \frac{h}{w}
          - \frac{125}{288} \frac{z}{w}
          + \frac{77}{18} \frac{w}{z}
          - \frac{45}{2} \frac{w}{(t-w)} \,,
\\
  x^{2,0}_{b,\mathrm{gl}} & =\frac{1}{w^2} \Bigg\{
             - \frac{9}{128} A_0(h)^2
             + A_0(h) A_0(t)   \left(
             - \frac{1}{8} \frac{t}{h}
             + \frac{1}{2} \frac{t}{h} N_c
             - \frac{3}{16} N_c
             - \frac{1}{32} \frac{h}{t}
             \right)
             \nonumber\\
           & + \frac{27}{64} h A_0(h) B_0(h;h,h)
             + A_0(h) B_0(h;t,t)   \left(
             - \frac{3}{4} \frac{t^2}{h} N_c
             + \frac{3}{16} t N_c
             \right)
             \nonumber\\
           & 
             + \frac{9}{64} hA_0(h) B_0(h;0,0)
             + A_0(h)   \left(
             \frac{31}{64} h
             - \frac{5}{8} \frac{t^2}{h} N_c
             + \frac{13}{128} t
             + \frac{3}{32} t N_c
             \right)
             \nonumber\\
           & 
             + A_{0,\varepsilon}(h)   \left(
             \frac{1}{8} \frac{t^2}{h}
             - \frac{9}{8} \frac{t^2}{h} N_c
             + \frac{3}{64} t
             + \frac{3}{8} t N_c
             + \frac{7}{8} h
             \right)
             \nonumber\\
           & + A_0^2(t)   \left(
             + \frac{39}{128}
             - \frac{1}{2} \frac{t^2}{h^2} N_c^2
             - \frac{3}{4} \frac{t}{h} N_c
             + \frac{1}{2} \frac{t}{h} N_c^2
             + \frac{1}{16} N_c
             - \frac{1}{32} \frac{h}{t}
             \right)
             \nonumber\\
           & 
             + \frac{9}{8} t N_cA_0(t) B_0(h;h,h) 
             + A_0(t) B_0(h;t,t)   \left(
             - 2 \frac{t^3}{h^2} N_c^2
             + \frac{1}{2} \frac{t^2}{h} N_c^2
             \right)
             \nonumber\\
           & 
             + \frac{3}{8} t N_c A_0(t) B_0(h;0,0)  
             + A_0(t) B_0(t;h,t)   \left(
             - 2 \frac{t^2}{h} N_c
             + \frac{3}{8} t
             + \frac{1}{2} t N_c
             - \frac{3}{32} h
             \right)
             \nonumber\\
           & 
             + A_0(t) B_0(t;0,0)   \left(
             - \frac{1}{2} \frac{t^2}{h} N_c
             + \frac{3}{32} t
             \right)
             \nonumber\\
           & + A_0(t)   \left(
             - \frac{25}{16} \frac{t^2}{h} N_c
             + \frac{1}{4} \frac{t^2}{h} N_c^2
             + \frac{85}{128} t
             - \frac{1}{16} t N_c
             + \frac{1}{64} h
             \right)
             \nonumber\\
           & 
             + A_{0,\varepsilon}(t)   \left(
             - \frac{11}{4} \frac{t^2}{h} N_c
             + \frac{5}{16} t
             + \frac{13}{16} t N_c
             + \frac{7}{64} h
             + \frac{1}{32} \frac{h^2}{t}
             \right)
             \nonumber\\
           & 
             + \frac{27}{64} h^2 B_0(h;h,h)  
             + B_0(h;t,t)   \left(
             - \frac{3}{4} t^2 N_c
             + \frac{3}{16} h t N_c
             \right)
             \nonumber\\
           & + \frac{9}{64} h^2 B_0(h;0,0) 
             + B_0(t;h,t)   \left(
             - \frac{t^3}{h} N_c
             + \frac{7}{16} t^2
             + \frac{1}{4} t^2 N_c
             - \frac{7}{64} h t
             \right)
             \nonumber\\
           & + B_0(t;0,0)   \left(
             - \frac{1}{4} \frac{t^3}{h} N_c
             + \frac{7}{64} t^2
             \right)
             - \frac{15}{64} h I_0(h,h,h)
             + I_0(h,t,t)   \left(
             \frac{9}{8} \frac{t^2}{h} N_c
             + \frac{1}{32} t \right.
             \nonumber\\
           & \left.- \frac{3}{8} t N_c
             - \frac{13}{64} h
             + \frac{1}{32} \frac{h^2}{t}
             \right)
             + I_0(0,h,t)   \left(
             - \frac{1}{8} \frac{t^2}{h}
             - \frac{5}{64} t
             + \frac{19}{64} h
             - \frac{3}{32} \frac{h^2}{t}
             \right)
             \nonumber\\
           & + I_0(0,0,h)   \left(
             - \frac{17}{64} h
             + \frac{1}{16} \frac{h^2}{t}
             \right)
             + I_0(0,0,t)   \left(
             \frac{1}{8} \frac{t^2}{h}
             + \frac{1}{2} \frac{t^2}{h} N_c
             - \frac{19}{64} t
             - \frac{1}{16} t N_c
             \right)
             \nonumber\\
           &  + \frac{3}{2} \frac{t^3}{h} N_c
             - \frac{293}{512} t^2
             + \frac{63}{128} t^2 N_c
             - \frac{13}{128} h t
             - \frac{3}{16} h t N_c
             - \frac{1}{2} h^2
             \Bigg\} \,.
\label{eq:x20b}
\end{align}
Equations~(\ref{eq:x10b})--(\ref{eq:x20b}) agree with Eqs.~(16)--(18) of
Ref.~\cite{Kniehl:2014yia}.
Notice that Eq.~(\ref{eq:x10b}) contains terms that behave as $1/b$ for
$M_b\to0$.
However, these singularities cancel rendering $x_b^{1,0}$ finite for $M_b=0$.
This may be seen by expanding the functions $B_0(b;z;b)$ and $B_0(b;w,t)$
through order $O(b)$, which yields
\begin{eqnarray}
  B_0(b;z,b) &=& \frac{A_0(b)-A_0(z)}{z} + \frac{b}{z} \left[ \frac{1}{2}
        - \frac{A_0(z)}{z} \right]  + O(b^2) \,,
\nonumber
\\
  B_0(b;w,t) &=& \frac{A_0(w)-A_0(t)}{t-w} + \frac{b}{t-w} \left[ \frac{1}{2}
        + \frac{A_0(w)+w}{t-w} + \frac{w(A_0(w)-A_0(t))}{(t-w)^2}\right]
 + O(b^2) \,.
\nonumber\\
&&
\end{eqnarray}


\boldmath
\section{Higgs-boson mass renormalization constant in the
$\overline{\mathrm{MS}}$ scheme}
\label{app:zmmH}
\unboldmath

The evaluations of the threshold corrections $\delta_x$ with $x=W,Z,H,t,b$ at
two loops, as described in this work, require the renormalization of the masses
$m_x$ at two loops.
In the $\overline{\mathrm{MS}}$ scheme, the bare masses $m_{0,x}$ are expressed
in terms of the renormalized masses $m_x(\mu)$ as
\begin{eqnarray}
   m_{0,B}^2 &=& m_B^2(\mu) Z_B(\mu)\,, \qquad B = W,Z,H\,,
\nonumber\\
   m_{0,f} &=& m_f(\mu) Z_f(\mu)\,, \qquad f = t,b\,,
\end{eqnarray}
where $Z_x$ are the mass renormalization constants.
The two-loop expressions for $Z_W$ and $Z_Z$ may be found in
Refs.~\cite{Jegerlehner:2001fb,Jegerlehner:2002em} and those for $Z_t$ and
$Z_b$ in Ref.~\cite{Kniehl:2014yia}.
In the remainder of this appendix, we present the last missing mass
renormalization constant, $Z_H$, in terms of
$\overline{\mathrm{MS}}$-renormalized parameters.
We have
\begin{eqnarray}
Z_{H} &=& 1
        + \frac{g^2}{16\pi^2}\, \frac{Z_{\alpha}^{(1,1)}}{\varepsilon}
        + C_F \frac{g_s^2}{16\pi^2}\,\frac{Z_{\alpha_s}^{(1,1)}}{\varepsilon}
        + C_F \frac{g^2}{16\pi^2}\,\frac{g_s^2}{16\pi^2}
           \left( \frac{Z_{\alpha\alpha_s}^{(2,2)}}{\varepsilon^2}
                  +\frac{Z_{\alpha\alpha_s}^{(2,1)}}{\varepsilon} \right)
\nonumber\\
&&{}    + \left( \frac{g^2}{16\pi^2} \right)^2
           \left( \frac{Z_{\alpha^2}^{(2,2)}}{\varepsilon^2}
                  +\frac{Z_{\alpha^2}^{(2,1)}}{\varepsilon} \right)
        + \cdots  \,,
\end{eqnarray}
where
\begin{eqnarray}
Z^{(1,1)}_{\alpha} &=&   N_c \frac{1}{2} \frac{m_t^2}{m_W^2}
          - \frac{3}{2}
          + \frac{3}{4} \frac{m_H^2}{m_W^2}
          - \frac{3}{4} \frac{m_Z^2}{m_W^2} \,,
\nonumber\\
Z^{(j,k)}_{\alpha_s} &=&   0 \,,
\nonumber\\
Z^{(2,2)}_{\alpha\alpha_s} &=& - N_c \frac{3}{2} \frac{m_t^2}{m_W^2} \,,
\nonumber\\
Z^{(2,1)}_{\alpha\alpha_s} &=&   N_c \frac{5}{4} \frac{m_t^2}{m_W^2} \,,
\nonumber\\
Z^{(2,2)}_{\alpha^2} &=& 
        \frac{41}{4}
          + \frac{27}{32} \frac{m_H^4}{m_W^4}
          - \frac{9}{8} \frac{m_Z^2 m_H^2}{m_W^4}
          + \frac{25}{32} \frac{m_Z^4}{m_W^4}
          - \frac{9}{4} \frac{m_H^2}{m_W^2}
          + \frac{5}{4} \frac{m_Z^2}{m_W^2}
       + n_G   \Bigg(
          - \frac{3}{4}
\nonumber\\
&&{}
          - \frac{3}{8} \frac{m_Z^4}{m_W^4}
          + \frac{3}{4} \frac{m_Z^2}{m_W^2}
          \Bigg)
       + N_c   \Bigg(
          - \frac{9}{16} \frac{m_t^4}{m_W^4}
          + \frac{3}{4} \frac{m_H^2 m_t^2}{m_W^4}
          - \frac{35}{48} \frac{m_Z^2 m_t^2}{m_W^4}
          - \frac{23}{24} \frac{m_t^2}{m_W^2}
          \Bigg)
\nonumber\\
&&{}
       + N_c n_G   \Bigg(
          - \frac{19}{36}
          - \frac{11}{72} \frac{m_Z^4}{m_W^4}
          + \frac{11}{36} \frac{m_Z^2}{m_W^2}
          \Bigg)
       + N_c^2 \frac{1}{4} \frac{m_t^4}{m_W^4} \,,
\nonumber\\
Z^{(2,1)}_{\alpha^2} &=& 
       - \frac{17}{3}
          - \frac{15}{64} \frac{m_H^4}{m_W^4}
          + \frac{3}{4} \frac{m_Z^2 m_H^2}{m_W^4}
          + \frac{157}{192} \frac{m_Z^4}{m_W^4}
          + \frac{3}{2} \frac{m_H^2}{m_W^2}
          - \frac{7}{6} \frac{m_Z^2}{m_W^2}
\nonumber\\
&&{}      + n_G   \Bigg(
            \frac{5}{8}
          + \frac{5}{16} \frac{m_Z^4}{m_W^4}
          - \frac{5}{8} \frac{m_Z^2}{m_W^2}
          \Bigg)
       + N_c   \Bigg(
          - \frac{9}{32} \frac{m_t^4}{m_W^4}
          - \frac{3}{8} \frac{m_H^2 m_t^2}{m_W^4}
\nonumber\\
&&{}      + \frac{85}{288} \frac{m_Z^2 m_t^2}{m_W^4}
          + \frac{25}{144} \frac{m_t^2}{m_W^2}
          \Bigg)
       + N_c n_G   \Bigg(
            \frac{95}{216}
          + \frac{55}{432} \frac{m_Z^4}{m_W^4}
          - \frac{55}{216} \frac{m_Z^2}{m_W^2}
          \Bigg) \,.
\end{eqnarray}

%
%


\begin{thebibliography}{99}

\bibitem{Aad:2012tfa}
  G.~Aad, et al., ATLAS Collaboration,
  Phys.\ Lett.\ B 716 (2012) 1,
  arXiv:1207.7214 [hep-ex];\\
  S.~Chatrchyan, et al., CMS Collaboration,
  Phys.\ Lett.\ B 716 (2012) 30,
  arXiv:1207.7235 [hep-ex].

\bibitem{Fischler:1982du}
  M.~Fischler, J.~Oliensis,
  Phys.\ Lett.\ B 119 (1982) 385;\\
  M.E.~Machacek, M.T.~Vaughn,
  Nucl.\ Phys.\ B 222 (1983) 83;\\
  M.E.~Machacek, M.T.~Vaughn,
  Nucl.\ Phys.\ B 236 (1984) 221;\\
  C.~Ford, I.~Jack, D.R.T.~Jones,
  Nucl.\ Phys.\ B 387 (1992) 373,
  arXiv:hep-ph/0111190;\\
  C.~Ford, I.~Jack, D.R.T.~Jones,
  Nucl.\ Phys.\ B 504 (1997) 551, Erratum;\\
  M.~Luo, Y.~Xiao,
  Phys.\ Rev.\ Lett.\ 90 (2003) 011601,
  arXiv:hep-ph/0207271.

\bibitem{Mihaila:2012fm}
  L.N.~Mihaila, J.~Salomon, M.~Steinhauser,
  Phys.\ Rev.\ Lett.\ 108 (2012) 151602,
  arXiv:1201.5868 [hep-ph];\\
  K.G.~Chetyrkin, M.F.~Zoller,
  J. High Energy Phys.\ 1206 (2012) 033,
  arXiv:1205.2892 [hep-ph];\\
  L.N.~Mihaila, J.~Salomon, M.~Steinhauser,
  Phys.\ Rev.\ D 86 (2012) 096008,
  arXiv:1208.3357 [hep-ph];\\
  A.V.~Bednyakov, A.F.~Pikelner, V.N.~Velizhanin,
  J. High Energy Phys.\ 1301 (2013) 017,
  arXiv:1210.6873 [hep-ph];\\
  A.V.~Bednyakov, A.F.~Pikelner, V.N.~Velizhanin,
  Phys.\ Lett.\ B 722 (2013) 336,
  arXiv:1212.6829 [hep-ph];\\
  K.G.~Chetyrkin, M.F.~Zoller,
  J. High Energy Phys.\ 1304 (2013) 091,
  arXiv:1303.2890 [hep-ph];\\
  K.G.~Chetyrkin, M.F.~Zoller,
  J. High Energy Phys.\ 1309 (2013) 155, Erratum;\\
  A.V.~Bednyakov, A.F.~Pikelner, V.N.~Velizhanin,
  Nucl.\ Phys.\ B 875 (2013) 552,
  arXiv:1303.4364 [hep-ph].

\bibitem{Jegerlehner:2001fb}
  F.~Jegerlehner, M.Yu.~Kalmykov, O.~Veretin,
  Nucl.\ Phys.\ B 641 (2002) 285,
  arXiv:hep-ph/0105304.

\bibitem{Jegerlehner:2002em}
  F.~Jegerlehner, M.Yu.~Kalmykov, O.~Veretin,
  Nucl.\ Phys.\ B 658 (2003) 49,
  arXiv:hep-ph/0212319.

\bibitem{Braaten:1980yq}
  E.~Braaten, J.P.~Leveille,
  Phys.\ Rev.\ D 22 (1980) 715;\\
  R.~Tarrach,
  Nucl.\ Phys.\ B 183 (1981) 384.

\bibitem{Gray:1990yh}
  N.~Gray, D.J.~Broadhurst, W.~Grafe, K.~Schilcher,
  Z.\ Phys.\ C 48 (1990) 673.

\bibitem{Fleischer:1998dw}
  J.~Fleischer, F.~Jegerlehner, O.V.~Tarasov, O.L.~Veretin,
  Nucl.\ Phys.\ B 539 (1999) 671,
  arXiv:hep-ph/9803493;\\
  J.~Fleischer, F.~Jegerlehner, O.V.~Tarasov, O.L.~Veretin,
  Nucl.\ Phys.\ B 571 (2000) 511, Erratum.

\bibitem{Chetyrkin:1999ys}
  K.G.~Chetyrkin, M.~Steinhauser,
  Phys.\ Rev.\ Lett.\ 83 (1999) 4001,
  arXiv:hep-ph/9907509;\\
  K.G.~Chetyrkin, M.~Steinhauser,
  Nucl.\ Phys.\ B 573 (2000) 617,
  arXiv:hep-ph/9911434;\\
  K.~Melnikov, T.~van~Ritbergen,
  Phys.\ Lett.\ B 482 (2000) 99,
  arXiv:hep-ph/9912391.

\bibitem{Marquard:2015qpa}
  P.~Marquard, A.V.~Smirnov, V.A.~Smirnov, M.~Steinhauser,
  Phys.\ Rev.\ Lett.\ 114 (2015) 142002,
  arXiv:1502.01030 [hep-ph].

\bibitem{Bednyakov:2002sf}
  A.~Bednyakov, A.~Onishchenko, V.~Velizhanin, O.~Veretin,
  Eur.\ Phys.\ J.\ C 29 (2003) 87,
  arXiv:hep-ph/0210258.

\bibitem{Hempfling:1994ar}
  R.~Hempfling, B.A.~Kniehl,
  Phys.\ Rev.\ D 51 (1995) 1386,
  arXiv:hep-ph/9408313.

\bibitem{Kniehl:2004hfa}
  B.A.~Kniehl, J.H.~Piclum, M.~Steinhauser,
  Nucl.\ Phys.\ B 695 (2004) 199,
  arXiv:hep-ph/0406254.

\bibitem{Jegerlehner:2003py}
  F.~Jegerlehner, M.Yu.~Kalmykov,
  Nucl.\ Phys.\ B 676 (2004) 365,
  arXiv:hep-ph/0308216.

\bibitem{Jegerlehner:2012kn}
  F.~Jegerlehner, M.Yu.~Kalmykov, B.A.~Kniehl,
  Phys.\ Lett.\ B 722 (2013) 123,
  arXiv:1212.4319 [hep-ph].

\bibitem{Bezrukov:2012sa}
  F.~Bezrukov, M.Yu.~Kalmykov, B.A.~Kniehl, M.~Shaposhnikov,
  J. High Energy Phys.\ 1210 (2012) 140,
  arXiv:1205.2893 [hep-ph].

\bibitem{Kniehl:2014yia}
  B.A.~Kniehl, O.L.~Veretin,
  Nucl.\ Phys.\ B 885 (2014) 459,
  arXiv:1401.1844 [hep-ph];\\
  B.A.~Kniehl, O.L.~Veretin,
  Nucl.\ Phys.\ B 894 (2015) 56, Erratum.

\bibitem{Sirlin:1985ux}
  A.~Sirlin, R.~Zucchini,
  Nucl.\ Phys.\ B 266 (1986) 389.

\bibitem{Degrassi:2012ry}
  G.~Degrassi, S.~Di Vita, J.~Elias-Mir\'o, J.R.~Espinosa, G.F.~Giudice,
  G.~Isidori, A.~Strumia,
  J. High Energy Phys.\ 1208 (2012) 098,
  arXiv:1205.6497 [hep-ph].

\bibitem{Buttazzo:2013uya}
  D.~Buttazzo, G.~Degrassi, P.P.~Giardino, G.~F.~Giudice, F.~Sala, A.~Salvio,
  A.~Strumia,
  J. High Energy Phys.\ 1312 (2013) 089,
  arXiv:1307.3536 [hep-ph].

\bibitem{Martin:2014cxa}
  S.P.~Martin, D.G.~Robertson,
  Phys.\ Rev.\ D 90 (2014) 073010,
  arXiv:1407.4336 [hep-ph].

\bibitem{Martin:2013gka}
  S.P.~Martin,
  Phys.\ Rev.\ D 89 (2014) 013003,
  arXiv:1310.7553 [hep-ph].

\bibitem{Sirlin:1980nh}
  A.~Sirlin,
  Phys.\ Rev.\ D 22 (1980) 971.

\bibitem{Malde:1999bd}
  P.~Malde, R.G.~Stuart,
  Nucl.\ Phys.\ B 552 (1999) 41,
  arXiv:hep-ph/9903403.

\bibitem{Awramik:2002vu}
  M.~Awramik, M.~Czakon, A.~Onishchenko, O.~Veretin,
  Phys.\ Rev.\ D 68 (2003) 053004,
  arXiv:hep-ph/0209084.

\bibitem{Onishchenko:2002ve}
  A.~Onishchenko, O.~Veretin,
  Phys.\ Lett.\ B 551 (2003) 111,
  arXiv:hep-ph/0209010.

\bibitem{Actis:2006rc} 
  S.~Actis, G.~Passarino,
  Nucl.\ Phys.\ B 777 (2007) 100,
  arXiv:hep-ph/0612124.

\bibitem{Fleischer:1980ub}
  J.~Fleischer, F.~Jegerlehner,
  Phys.\ Rev.\ D 23 (1981) 2001.

\bibitem{Kniehl:2012zb}
  B.A.~Kniehl and A.~Sirlin,
  Phys.\ Rev.\ D 85 (2012) 036007,
  arXiv:1201.4333 [hep-ph];\\
  B.A.~Kniehl,
  Phys.\ Rev.\ Lett.\ 112 (2014) 071603,
  arXiv:1308.3140 [hep-ph];\\
  B.A.~Kniehl,
  Phys.\ Rev.\ D 89 (2014) 096005;\\
  B.A.~Kniehl,
  Phys.\ Rev.\ D 89 (2014) 116010,
  arXiv:1404.5908 [hep-th].

\bibitem{Jegerlehner:2002er}
  F.~Jegerlehner, M.Yu.~Kalmykov, O.~Veretin,
  Nucl.\ Phys.\ B (Proc.\ Suppl.) 116 (2003) 382,
  arXiv:hep-ph/0212003.

\bibitem{Degrassi:1990tu}
  G.~Degrassi, S.~Fanchiotti, A.~Sirlin,
  Nucl.\ Phys.\ B 351 (1991) 49.

\bibitem{Fanchiotti:1992tu}
  S.~Fanchiotti, B.A.~Kniehl, A.~Sirlin,
  Phys.\ Rev.\ D 48 (1993) 307,
  arXiv:hep-ph/9212285.

\bibitem{Sturm:2013uka}
  C.~Sturm,
  Nucl.\ Phys.\ B 874 (2013) 698,
  arXiv:1305.0581 [hep-ph].

\bibitem{Erler:1998sy}
  J.~Erler,
  Phys.\ Rev.\ D 59 (1999) 054008,
  arXiv:hep-ph/9803453.

\bibitem{Agashe:2014kda} 
  K.A.~Olive, et al., Particle Data Group Collaboration,
  Chin.\ Phys.\ C 38 (2014) 090001.

\bibitem{Baikov:2012zm}
  P.A.~Baikov, K.G.~Chetyrkin, J.H.~K\"uhn, J.~Rittinger,
  J. High Energy Phys.\ 1207 (2012) 017,
  arXiv:1206.1284 [hep-ph];\\
  P.A.~Baikov, K.G.~Chetyrkin, J.H.~K\"uhn, C.~Sturm,
  Nucl.\ Phys.\ B 867 (2013) 182,
  arXiv:1207.2199 [hep-ph].

\bibitem{Schroder:2005hy}
  Y.~Schr\"oder, M.~Steinhauser,
  J. High Energy Phys.\ 0601 (2006) 051,
  arXiv:hep-ph/0512058;\\
  K.G.~Chetyrkin, J.H.~K\"uhn, C.~Sturm,
  Nucl.\ Phys.\ B 744 (2006) 121,
  arXiv:hep-ph/0512060;\\
  B.A.~Kniehl, A.V.~Kotikov, A.I.~Onishchenko, O.L.~Veretin,
  Phys.\ Rev.\ Lett.\ 97 (2006) 042001,
  arXiv:hep-ph/0607202.

\bibitem{cpc}
B.A.~Kniehl, A.F.~Pikelner, and O.L.~Veretin, in preparation.

\bibitem{Martin:2005qm}
  S.P.~Martin, D.G.~Robertson,
  Comput.\ Phys.\ Commun.\ 174 (2006) 133,
  arXiv:hep-ph/0501132.

\bibitem{Tarasov:1996br}
  O.V.~Tarasov,
  Phys.\ Rev.\ D 54 (1996) 6479,
  arXiv:hep-th/9606018.

\bibitem{Tarasov:1997kx}
  O.V.~Tarasov,
  Nucl.\ Phys.\ B 502 (1997) 455,
  arXiv:hep-ph/9703319.

\end{thebibliography}
\end{document}